\documentclass[a4paper,fleqn,usenatbib]{mnras}

\usepackage[T1]{fontenc}
\usepackage{ae,aecompl}

\usepackage{graphicx}	
\usepackage{amsmath}	
\usepackage{amssymb}	
\usepackage{float}
\usepackage{dblfloatfix}
\usepackage{subfigure}
\usepackage{blindtext}
\usepackage[dvipsnames]{xcolor}
\usepackage{enumitem}
\usepackage[colorinlistoftodos]{todonotes}

\usepackage{scalerel}
\usepackage{tikz}
\usetikzlibrary{svg.path}

\definecolor{orcidlogocol}{HTML}{A6CE39}
\tikzset{
  orcidlogo/.pic={
    \fill[orcidlogocol] svg{M256,128c0,70.7-57.3,128-128,128C57.3,256,0,198.7,0,128C0,57.3,57.3,0,128,0C198.7,0,256,57.3,256,128z};
    \fill[white] svg{M86.3,186.2H70.9V79.1h15.4v48.4V186.2z}
                 svg{M108.9,79.1h41.6c39.6,0,57,28.3,57,53.6c0,27.5-21.5,53.6-56.8,53.6h-41.8V79.1z M124.3,172.4h24.5c34.9,0,42.9-26.5,42.9-39.7c0-21.5-13.7-39.7-43.7-39.7h-23.7V172.4z}
                 svg{M88.7,56.8c0,5.5-4.5,10.1-10.1,10.1c-5.6,0-10.1-4.6-10.1-10.1c0-5.6,4.5-10.1,10.1-10.1C84.2,46.7,88.7,51.3,88.7,56.8z};
  }
}

\newcommand\orcidicon[1]{\href{https://orcid.org/#1}{\mbox{\scalerel*{
\begin{tikzpicture}[yscale=-1,transform shape]
\pic{orcidlogo};
\end{tikzpicture}
}{|}}}}

\newcommand{\eagle}{{\sc eagle}}
\newcommand{\ikea}{\hbox{I$\kappa\epsilon\alpha$}}

\title[Mass, metallicity \& morphology in \eagle]
{The origin of correlations between mass, metallicity and morphology in galaxies from the \eagle\ simulation}

\author[Zenocratti et al.]{
L. J. Zenocratti$^{1,2,\orcidicon{0000-0001-8271-1794}}$\thanks{E-mail: ljzenocratti@gmail.com},
M. E. De Rossi$^{3,4,\orcidicon{0000-0002-4575-6886}}$\thanks{E-mail: mariaemilia.dr@gmail.com},
T. Theuns$^{5,6,\orcidicon{0000-0002-3790-9520}}$,
M. A. Lara-L\'opez$^{7,8,9,\orcidicon{0000-0001-7327-3489}}$
\\
$^{1}$Facultad de Ciencias Astron\'omicas y Geof\'isicas, Universidad Nacional de La Plata, Paseo del Bosque s/n, La Plata, Argentina\\
$^{2}$Instituto de Astrofísica de La Plata, CONICET-Universidad Nacional de La Plata, Paseo del Bosque s/n, La Plata, Argentina\\
$^{3}$Universidad de Buenos Aires, Facultad de Ciencias Exactas y Naturales y Ciclo B\'asico Com\'un, Buenos Aires, Argentina\\ 
$^{4}$CONICET-Universidad de Buenos Aires, Instituto de Astronom\'ia y F\'isica del Espacio (IAFE), Buenos Aires, Argentina\\
$^{5}$Institute for Computational Cosmology, Durham University, South Road, Durham, DH1 3LE, UK\\
$^{6}$Physics Department, Durham University, South Road, Durham, DH1 3LE, UK\\
$^{7}$Armagh Observatory and Planetarium, College Hill, Armagh BT61 9DG, Northern Ireland, UK\\
$^{8}$Departamento de Física de la Tierra y Astrofísica, Universidad Complutense de Madrid, E-28040 Madrid, Spain\\
$^{9}$Instituto de Física de Partículas y del Cosmos IPARCOS, Fac. de Ciencias Físicas, Universidad Complutense de Madrid, \\ E-28040, Madrid, Spain\\
}

\date{Accepted XXX. Received YYY; in original form ZZZ}

\pubyear{2021}

\begin{document}
\label{firstpage}
\pagerange{\pageref{firstpage}--\pageref{lastpage}}
\maketitle

\begin{abstract}

Observed and simulated galaxies exhibit correlations between stellar mass, metallicity and morphology. We use the \eagle\ cosmological simulation to examine the origin of these correlations for galaxies in the stellar mass range $10^9~\rm{M_\odot} \leqslant\ M_\star \leqslant 10^{10}~\rm{M_\odot}$, and the extent to which they contribute to the scatter in the mass-metallicity relation. We find that rotationally supported disc galaxies have lower metallicity than dispersion supported spheroidal galaxies at a given mass, in agreement with previous findings. In \eagle\, this correlation arises because discs form stars at later times, redshift $z\leqslant 1$, from the accretion of low-metallicity gas, whereas spheroidal galaxies galaxies typically form stars earlier, mainly by consumption of their gas reservoir. The different behaviour reflects the growth of their host dark matter halo: at a given stellar mass, disc galaxies inhabit dark matter haloes with lower mass that formed later compared to the haloes of spheroidal galaxies. Halo concentration plays a secondary role.

\end{abstract}




\section{Introduction}
\label{sec:intro}

The gas-phase metallicity, $Z$, of galaxies encodes information on how physical processes drive the evolution of galaxies (see recent discussions in e.g. \citealp{Dave2012}; \citealp{Finlator2017}). The emergence of scaling relations between $Z$ and other galaxy properties such as stellar mass, $M_\star$, gas fraction and stellar age, plays a fundamental role in understanding galaxy formation processes and constraining physical models of galaxy evolution. The relation between $Z$ and $M_\star$,
the mass-metallicity relation (hereafter MZR), has received particular attention 
in the past few years.

The observed MZR extends over several orders of magnitude in $M_\star$ in the local Universe, with more massive galaxies more metal-enriched following a power-law relation
$Z\propto M_\star^\alpha$, with slope $\alpha\approx 0.4$ for $M_\star\leqslant 10^{10}M_\odot$
and a flatter or even inverted relation at higher mass \cite[e.g.][]{Lequeux1979, Tremonti2004}. A MZR relation is also detected at higher redshifts, $z$, although
the slope and normalization may evolve \cite[e.g.][]{Maiolino2008, Troncoso2014}. However, a detailed comparison between the low and high-$z$ MZR is complicated by issues of sample bias and the use of different metallicity indicators at different $z$, amongst other challenges \cite[e.g.][]{Kewley2008, Steidel2014, Telford2016}.

The scatter of galaxies around the median MZR correlates with other observables. For example, \citet{Ellison2008} showed that galaxies with lower star formation rate, SFR, or smaller half-mass radius, have higher $Z$ at given $M_\star$. \citet{Laralopez2010} and \citet{Mannucci2010} suggested that this may be an indication that
the MZR relation is the 2D projection of a more fundamental underlying relationship,
for example between $Z$, $M_\star$ and SFR, which was dubbed
the \lq fundamental metallicity relation\rq, hereafter FMR (see \citealp{Curti2020} for a recent review on the MZR and its relation to the FMR). Alternatively, the MZR could
emerge from a relation between $M_\star$, $Z$ and the gas fraction, $f_{\rm g}$, since
galaxies with higher $f_{\rm g}$ tend to exhibit lower $Z$ at a given $M_\star$
\cite[e.g.][]{Bothwell2013, Laralopez2013a}. More recently, \citet{Wu2019} showed
that observed galaxies with higher concentration index, higher Sérsic index or higher SFR, tend to be more metal-poor at given $M_\star$. Although similar findings at different $z$ have been reported in the literature \cite[e.g.][]{Hunt2012, Laralopez2013b, Cullen2014, Zahid2014, Bothwell2016}, uncertainties resulting from sample selection and the usage of different metallicity indicators have hampered attempts at building a unified view of what the observations imply.
 
The median MZR and its scatter have also been studied theoretically. 
The semi-analytical model by \citet{Calura2009} reproduces the observed correlation between $Z$, $M_\star$ and SFR by positing that the efficiency of star formation increases with $M_\star$, independent of galaxy morphology. \citet{Dave2012} describes an analytical model of the baryon cycle in galaxies in terms of a slowly evolving equilibrium state between accretion, star formation and outflows powered by stellar feedback. In this formalism, $Z\propto {\rm SFR}$, with the proportionality factor proportional to the yield and inversely proportional to the inflow rate. In the models by \citet{Lilly2013} and \citet{Forbes2014}, the metallicity of a galaxy approaches an equilibrium value on the gas consumption time-scale, $\tau_{\rm gas}$. These authors argue that $\tau_{\rm gas}$ is short and, consequently, $Z$ is close to the equilibrium metallicity, which depends on 
the yield and, inversely, on the specific star formation rate, sSFR=SFR/$M_\star$.
In the $I\kappa \epsilon \alpha$ model of self-regulated galaxy formation by
\citet{Sharma2020}, the metallicity approaches an equilibrium value which depends on $v^2_h$, a measure of host halo's potential. The power-law MZR relation $Z \propto M_\star^{2/5}$ emerges because the star formation efficiency depends on $v^2_h$. \citet{Fontanot2021} studied the evolution of the MZR and the FMR as predicted by the GAlaxy Evolution and Assembly ({\sc{gaea}}) semi-analytic model, comparing them with recent results from the {\sc vandels} survey (\citealp{Mclure2018}; \citealp{Pentericci2018}).

Cosmological hydrodynamical simulations that include stellar evolution and enrichment also produce a MZR with a slope and normalization that depend somewhat on the parameters of the sub-grid model. \citet{Lagos2016} showed that the correlations that give rise to the FMR in the \eagle\ simulations (\citealp{Schaye2015, Crain2015}) are a consequence of galaxies lying on a two-dimensional plane in the space of gas fraction, star formation rate and stellar mass, which they dubbed the \lq Fundamental Plane\rq. They argue that this plane is a consequence of self-regulation. \citet{DeRossi2017} analysed the evolution of the MZR of galaxies in \eagle, concluding that the simulated galaxies follow the observed FMR well (see also \citealp{Laralopez2019}). In particular, the metallicity in terms of the O/H abundance of \eagle\ galaxies anti-correlates with the SFR at low $M_\star$, and correlates with SFR at high mass. \cite{Furlong2017} showed that galaxy sizes correlate with sSFR in \eagle, with star-forming galaxies typically larger than passive galaxies at a given stellar mass, as also seen in the data. Related to this, \citet{Sanchezalmeida2018} showed that the \eagle\ simulations also reproduce the observed metallicity dependence on galaxy sizes. \citet{Torrey2019} showed that galaxies identified in the {\sc{IllustrisTNG}} cosmological simulation (\citealp{Weinberger2017}; \citealp{Pillepich2018}) broadly reproduce the evolution of the observed MZR at $10^9 < M_\star/M_\odot < 10^{10.5}$ in the redshift interval $0 < z < 2$, and that the scatter of the simulated MZR correlates with gas mass and SFR, in agreement with the observed FMR. \citet{Dave2019, Dave2020} reported that galaxies in their {{\sc simba}} hydrodynamical simulations show correlations between $M_\star$, $Z$, sSFR and galaxy size, that agree quite well with observed trends, as well as with the results obtained from 
\eagle\ and {\sc{IllustrisTNG}}. Finally, \citet{VanLoon2021} studied the relation between accretion and outflows, and the scatter in the MZR in \eagle\ galaxies, showing that, at low stellar masses, there are negative correlations between the residuals of the MZR and the residuals of the relations between stellar mass and specific inflow and outflow rates, while, at high stellar masses, the correlations between residual metallicity and residual specific inflow and outflow rates are positive for relatively low flow rates, and negative for relatively high flow rates.

\citet{Zenocratti2020} showed that the scatter in the MZR of \eagle\ galaxies correlates with morphology and kinematics; a similar correlation is also seen in the {\sc{IllustrisTNG}} simulations. Such trends were not previously discussed: they might offer insight in the origin of the FMR and its scatter. According to \citet{Zenocratti2020}, at a given $M_{\star}$, $z=0$ star-forming \eagle\ galaxies with low metallicity tend to be flattened and rotation supported, whereas those at high $Z$ tend to be dispersion supported and spheroidal in shape. In short: galaxies with low-$Z$ for their stellar mass are discs, whereas those with high-$Z$ are early type galaxies. This holds true in the mass range $10^9\leqslant M_{\star}/{\rm M}_\odot\leqslant 10^{10}$. With increasing $M_\star$, $Z$ of disc galaxies increases while that of the early types remains approximately constant.
As a consequence, eventually $Z$ tends to be {\em lower} for early type galaxies compared to discs. The transition mass is around $10^{10}~\rm{M}_\odot$.
Above this characteristic mass, the typical $Z$ of early type galaxies {\em decreases} with $M_{\star}$, presumably because their AGN quenches star formation so that these galaxies can only grow in mass by mergers - typically with lower mass hence lower $Z$ galaxies. These trends are seen not just at $z=0$ but are present in \eagle\ at least up to $z\sim 3$.

In the case of $M_{\star}\leqslant 10^{10}\,{\rm M}_\odot$ star-forming galaxies, the recent accretion of metal-poor gas has been suggested as a driver of the secondary dependence of $Z$ on SFR, gas fraction and galaxy sizes (e.g. \citealp{DeRossi2017}; \citealp{Sanchezalmeida2018}; \citealp{Torrey2019}; \citealp{DeLucia2020}; \citealp{Wang2021}): the inflow increases the SFR and galaxy size, while decreasing $Z$ \citep{Zenocratti2020}. And, in a discy geometry, metal-loaded galactic outflows could be more efficient, leading to a decrease of the galaxy metallicity and gas fraction \citep{Creasey2013}. 
It is also worth mentioning that  cosmological gas accretion through cold streams has been proposed as a main mode of formation of disc and spheroids in the young Universe, alternative to mergers events \citep[e.g.][]{Dekel2009}.
The goal of the current paper is to extend the study carried out by \citet{Zenocratti2020}, examining in more detail the origins of these correlations and the impact on the MZR at $10^9~\rm{M_\odot} \leqslant\ M_\star \leqslant 10^{10}~\rm{M}_\odot$.

The structure of this paper is as follows. In Section \ref{sec2}, we briefly describe the \eagle\ suite and the selection of the galaxy sample. We also introduce the morpho-kinematical parameters used to characterize simulated galaxies and describe how we quantify inflows and outflows. In Section \ref{sect:expections}, we state some simple predictions regarding the MZR and its scatter. In Section \ref{sec:results}, we show our results regarding the $z=0$ MZR and the evolutionary histories of galaxies in our sample. In Section \ref{sec:discussion}, we discuss our results and compare to previous works. Finally, in Section \ref{sec:summary} we summarize our main findings. A convergence test using a higher-resolution simulation of the \eagle\ suite is presented in Appendix \ref{convergence}.

\begin{table*}
\centering
\caption{Box sizes and particle resolution of the main \eagle\ simulations used in this work. From left to right, the columns show the simulation name, comoving box size, particle number per species (i.e. gas, DM), initial baryonic particle mass, dark matter particle mass, comoving gravitational softening length, and maximum proper softening length. Note that different units of length are used (e.g., proper kiloparsec, denoted by pkpc, and comoving megaparsec, denoted by cMpc).
The masses of star particles are typically $\approx m_g$, with any difference a result of stellar mass-loss and enrichment.}
\begin{tabular*}{0.7\textwidth}{@{\extracolsep{\fill}} ccccccc}
\hline
Name & $L~\rm{[cMpc]}$ & $N$ & $m_{\rm g}~[\rm{M}_\odot]$ & $m_{\rm DM}~[\rm{M}_\odot]$ & $\epsilon_{\rm com}~[\rm{ckpc}]$ & 
$\epsilon_{\rm prop}~[\rm{pkpc}]$ \\
\hline
L0025N0752 & 25 & $752^3$ & $2.26 \times 10^5$ & $1.21 \times 10^6$ & 1.33 & 0.35 \\
L0100N1504 & 100 & $1504^3$ & $1.81 \times 10^6$ & $9.70 \times 10^6$ & 2.66 & 0.70 \\
\hline
\end{tabular*} 
\label{Tab1}
\end{table*}

\section{Simulations and methods} 
\label{sec2}

\subsection{The EAGLE Simulations}
\label{sec:simulation}

\eagle\ (Evolution and Assembly of GaLaxies and their Environment, \citealp{Schaye2015}) is a suite of cosmological hydrodynamical simulations
in a $\Lambda$CDM cosmology, performed with a modified version of the {\sc treePM-SPH} {\sc gadget 3} code (\citealp{Springel2005}). The parameters of the sub-grid physics modules that account for unresolved physical processes, were calibrated so that the simulations reproduce the observed $z\approx0$ galaxy stellar mass function, stellar mass - size relation, and the black hole mass - stellar mass relation, as described by \cite{Crain2015}. These sub-grid modules include prescriptions for star formation and for the feedback associated with massive stars. The model also includes stellar evolution and the synthesis
and dispersal of metals produced in stars, in particular tracking
11 chemical elements (H, He, C, N, O, Ne, Mg, Si, S, Ca, and Fe), as described
by \cite{Wiersma09a}; these same elements affect the cooling and photo-heating of gas as described by \cite{Wiersma09b}. The suite also includes a model for the seeding, growth, and merging of supermassive black holes, and their impact as active galactic nuclei on their surroundings; see \cite{Schaye2015} for full details.

The adopted cosmological parameters are taken from \cite{Planck2015}, briefly $\Omega_{\Lambda}=0.693$, $\Omega_{\rm{m}}=0.307$, $\Omega_{\rm{b}}=0.04825$, $n_{\rm{s}}=0.9611$, $Y=0.248$, and $h=0.677$, where symbols have their usual meaning. The initial conditions consisted in a glass-like particle initial configuration, perturbed according to second-order Lagrangian perturbation theory using the method of \citet{Jenkins2010} and the public Panphasia Gaussian white noise field (\citealp{Jenkins2013}); details about the generation of the initial conditions can be found in Appendix B of \cite{Schaye2015}.

The \eagle\ suite includes runs in which the linear co-moving extent L of the simulation volume is varied, and in which the particle count, N$^3$ is varied.
L and N  are used to identify a given simulation, for example L0100N1504 is an \eagle\ simulation in a volume with linear extent 100 co-moving megaparsecs (cMpc), performed with 1504$^3$ gas and an equal number of dark matter particles. Some relevant parameters for the runs used in this work are summarized in Table \ref{Tab1}. Particle properties for all \eagle\ models were recorded for 29 snapshots between redshifts 20 and 0 (snapshots 0 and 28, respectively). We will refer to simulations with the same mass and spatial resolution as L0100N1504 
as \lq intermediate-resolution simulations\rq, and those with the same resolution as L0025N0752 as \lq high-resolution simulations\rq. Simulations that adopt the same sub-grid parameters as L0100N1504 are referred to as reference models. There is a strand of high-resolution \eagle\ simulations in which the sub-grid parameters were varied to improve the fit with the $z=0$ reference data, such as RecalL0025N0752. 

Dark matter haloes are identified using the friends-of-friends algorithm (FoF, \citealp{Davis1985}), whereas galaxies are identified using
the {\sc subfind} algorithm \citep{Springel2005, Dolag2009}. The galaxy 
that contains the most bound particle in a halo is identified with the central galaxy in that halo, while the remaining galaxies are then classified as satellites. 
A merger tree links a galaxy to its progenitors. Properties of galaxies, haloes and merger trees can be queried in the public \eagle\ database described by \cite{Mcalpine2016}. The \eagle\ particle data is also available as described by \cite{eagle2017}.

\subsection{Galaxy sample and morpho-kinematical parameters}
\label{sec:galaxy_sample}

In this paper, we use mainly simulation L0100N1504 because of its superior statistics compared to the smaller volumes. We note that the slope and normalization of the MZR in L0025N0752 agrees better with observations \citep{Schaye2015, DeRossi2017}, however investigating the origin of secondary correlations benefits greatly from the increased statistics provided by L0100N1504 \citep{DeRossi2017, Zenocratti2020}. The database for L0100N1504 includes entries for the metallicity, $Z$, of stars and star-forming gas, the gas fraction, $f_{\rm g}$, the sSFR and
the kinematic parameter $\kappa_{\rm co}$, which is the ratio of kinetic energy in rotation compared to the total kinetic energy in the centre of mass rest frame of each galaxy. In Appendix~\ref{convergence}, we demonstrate that the main results and conclusions as derived from L0100N1504 are consistent with those from Recal-L0025N0752.

To mitigate the impact of poorer resolution on our results, we restrict our sample to $z=0$ galaxies with $M_\star \geqslant 10^9~\rm{M}_\odot$. In order to mimic the aperture of instruments used for observations (see \citealt{Schaye2015}, for a discussion), we follow \citet{DeRossi2017} and \citet{Zenocratti2020},  and 
measure baryonic properties within spherical apertures of 30 proper kilo-parsecs (pkpc). Nevertheless, considering our upper stellar mass limit ($M_\star \lesssim 10^{10}~\rm{M}_\odot$), aperture effects are expected to be negligible \citep[e.g.][]{Schaye2015, DeRossi2017}.
We characterize the \lq metallicity\rq\ of star-forming (SF) gas by its O/H abundance; hence, the MZR studied in this work corresponds to the relation between $M_\star$ and O/H abundance of SF gas. This makes a comparison with observational metallicities derived from H{\sc II} regions more straightforward. In addition to a cut in stellar mass, we restrict 
our analysis sample to central{\footnote{Our main results remain unchanged if we also include satellite galaxies, but they are excluded in this work because of the environmental effects to which they are subjected. We leave the analysis of the effects of environment on our findings for a future work.}} galaxies with at least 25 SF gas particles (gas mass of, at least, $5.25\times 10^7~{\rm M}_\odot$). We further adopt an {\em upper} mass limit of $M_\star = 10^{10}~\rm{M}_\odot$. Above this value, the trends in secondary metallicity dependences inverts \citep{DeRossi2017, Zenocratti2020}, and in this work we investigate the physics that drives the trends below this characteristic stellar mass.

We use $\kappa_{\rm co}$ to distinguish between rotation- and dispersion-supported galaxies, and the disc-to-total stellar mass ratio, $D/T$, to classify galaxies as either disc or spheroid. In addition, we use the ellipticity ($\epsilon_\star$) of the stellar body, and its triaxiality ($T_\star$). These parameters were computed by \citet{Thob2019} for galaxies with at least 300 star particles within 30~pkpc of the centre of the potential of each galaxy. This results in a sample of 4470 galaxies, when all selections are accounted for.

The \eagle\ database includes values of $D/T$ as computed by \citet{Thob2019}\footnote{The $D/T$ ratios available in \eagle\ database were calculated indirectly as $D/T=1-B/T=1-2\frac{1}{M_\star}\sum\limits_{{\rm{i}},L_{\rm{z,i}}<0}m_{\rm{i}}$, where the sum is over all counter-rotating stellar particles within 30 pkpc, $m_{\rm{i}}$ is the mass of each stellar particle, and $L_{\rm{z,i}}$ is the component of its angular momentum projected along the direction of the total angular momentum vector of all stellar particles within the mentioned spherical radius.}. However, we performed our own calculations of $D/T$ with a different method that allows us to assign individual particles to the disc or spheroid component.
This classification allows for a more detailed correlation between kinematics and morphology of individual components. We assume that star particles inside 
30~pkpc with $j_{\rm{z}}/ j > 0.7$ are disc particles and the remainder are spheroid particles. Here, $j$ is the magnitude of the total angular momentum of the particle, and $j_{\rm{z}}$ the component along the $z$-axis. The latter is defined by the total angular momentum of all star particles. Values of $D/T$ computed using this criterion correlate very well
with those derived by \cite{Thob2019}.

\subsection{Inflows, outflows and particle tracking}
\label{sec:accretion}

To analyse the effects of mass accretion for simulated galaxies in our sample, we build the main branch of each galaxy's merger tree and compare the baryonic particles associated to the main progenitor between all successive two snapshots of the simulation\footnote{Data for re-constructing the merger trees were extracted from the \eagle\ database; see \citealt{Mcalpine2016}, for details.}. As mentioned before, we use a spherical aperture of 30 pkpc to compute properties of every progenitor and identify the particles that build it up. Basically, for a given merger tree, we consider the gas and star particles within every main progenitor, and determine which particles present in the galaxy at the snapshot $k+1$ are not present in the system at the snapshot $k$; those particles are then considered {\it accreted particles}. 
Our algorithm also tracks the conversion between different types of particles (i.e. the formation of a star particle from a SF gas particle, or the transformation of a SF gas particle in a non-SF gas particle and vice versa) between consecutive snapshots. And, for the selected accreted and {\it transformed} particles, we calculate their properties, such as their masses, nuclei abundances, and positions and velocities with respect to the centre of potential of the corresponding galaxy. From now on, in this work, we will use the term \lq {\em inflow}\rq\ to denote both, accreted and transformed particles, that become part of a given component (i.e stellar phase, SF gas or non-SF gas) of the galaxy between two successive snapshots of the simulation (for example, at a given snapshot, \lq inflow of SF gas\rq\ includes SF gas particles that were not inside the galaxy at the previous snapshot, or that were non-SF (NSF) gas particles at the previous snapshot and are SF gas particles in the galaxy at the current snapshot).

In a similar way, we quantify the \lq {\em outflow}\rq\ of particles. For a given merger tree, we determine which particles are present inside our default spherical aperture at the snapshot $k$, but not in the snapshot $k+1$, (irrespective of the type of transformation this particle might undergo). Those particles are considered {\it outflowing} particles and we classify them according to the galaxy component (i.e stellar phase, SF gas or non-SF gas) to which they belong at snapshot $k$. 

Computing the rate at which gas particles change from being non-star forming to star forming as a way to determine the {\em inflow} rate is clearly an oversimplification. It would be more accurate to track particles back in time to infer whether they are inflowing or outflowing, as was done by \cite{VanLoon2021} using the method discussed by \cite{Mitchell2020}. However, \cite{Mitchell2020} find that the majority of gas being accreted on to galaxies in \eagle\ is infalling for the first time, which also implies that our approximate inflow rate is presumably very similar to that used by \cite{VanLoon2021}.

The main branch of the merger tree of a given galaxy is followed back in time until we reach the oldest progenitor which has its morpho-kinematics parameters calculated according to \cite{Thob2019}. We consider that this first progenitor is the first \lq well-defined\rq\ galaxy of the corresponding tree, in the sense that it has an adequate number of particles (at least 300 star particles) to characterize its morphology. 
 
Although the study of inflow and outflow of SF gas particles will be presented in Sec. \ref{sec:average_inout}, we define here some quantities required for such analysis. As mentioned before, an \lq inflow of SF gas\rq\ is given by particles 
that, at a given snapshot in the simulation, are SF gas particles but were NSF gas particles or were outside the galaxy at the previous snapshot. Thus, $M_{\rm inflow}$, $\dot{M}_{\rm inflow}$, and $\rm (O/H)_{\rm inflow}$ depict the mass (strictly speaking, the sum of individual masses), the inflow rate, and 
the average O/H abundance of the inflowing SF gas particles in a galaxy at a given redshift, respectively. Each inflowing SF gas particle in the galaxy 
has an associated O/H abundance at a given snapshot that can be compare with the overall abundance of the galaxy at the previous snapshot; 
then, we define the (mass) fraction of metal-poor inflowing SF gas ($F_{\rm inflow,\ low}$) as the fraction of inflowing SF gas particles whose O/H, at a given snapshot, 
is lower than the average O/H of the galaxy at the previous snapshot. Similar notation is applied for quantities associated to \lq outflows\rq. Finally, we note that our analysis uses the \eagle\ snapshots, which sample the galaxies at intervals separated by typically $\sim 1$~Gyr below redshift $z=1$, and shorter above that redshift \citep{Mcalpine2016}. The output frequency of snapshots determines the shortest timescales that we can sample.

\section{Expectations for the median MZR and its scatter}
\label{sect:expections}
\begin{figure}
\centering
\subfigure{\includegraphics[width=0.9\columnwidth]{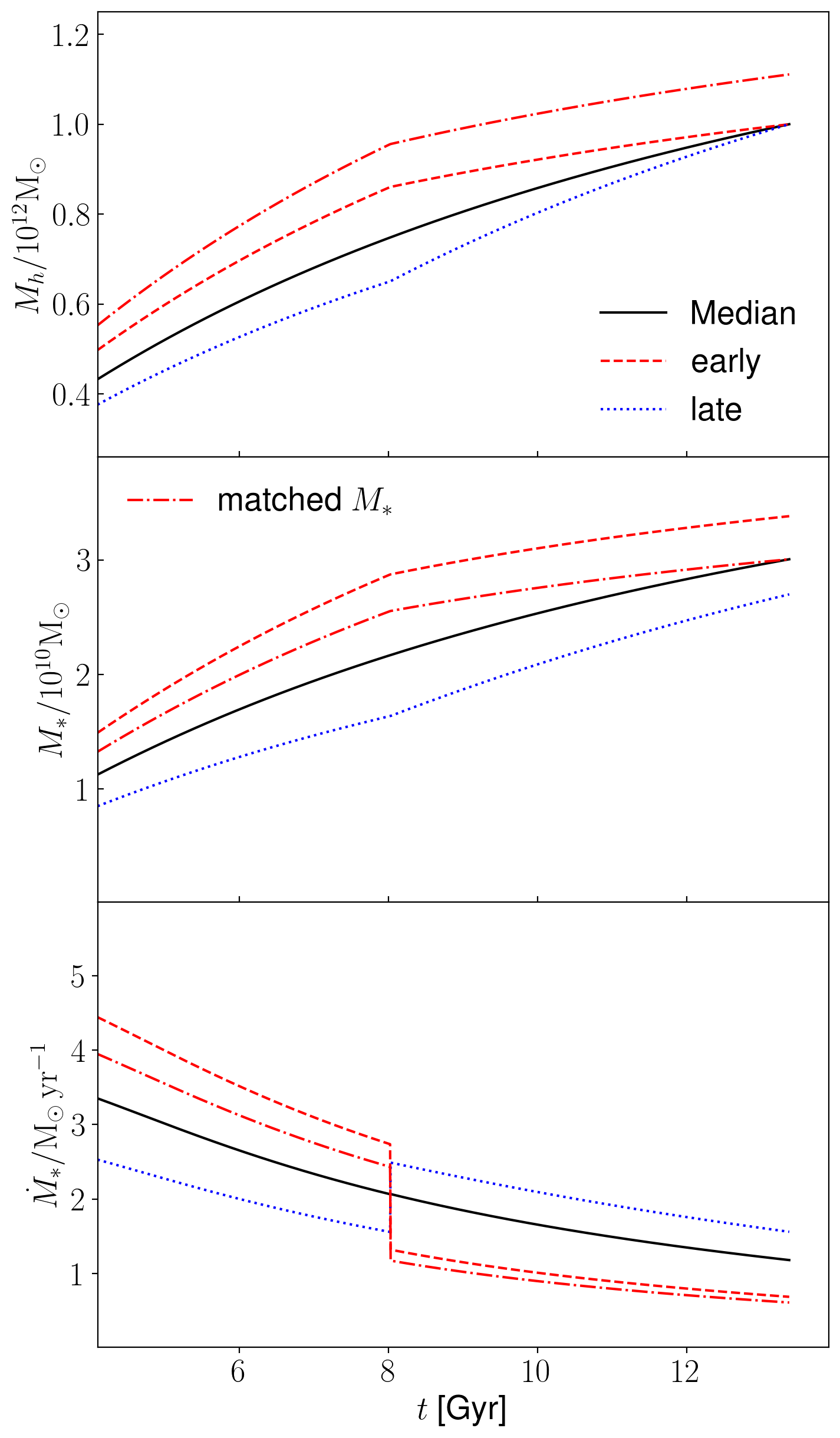}}
\caption{Cartoon illustrating the effect of the formation history and concentration parameter on halo mass ({\em upper panel}), stellar mass ({\em central panel}) and star formation rate ({\em lower panel}); the latter two are computed using the \ikea\ model of \protect\cite{Sharma2020}. The {\em solid black line} uses the mean halo accretion history of \protect\cite{Correa2017}; the {\em dashed red} halo grows faster initially and slower later on, so that its final mass is the same, as may be the case for a halo with higher than average concentration (and vice versa for the {\em dotted blue} halo). The halo in red hosts a higher mass galaxy than average whose current star formation rate is lower than average (and vice versa for the lower concentration blue halo). The {\em dot-dashed red line} is for a higher mass halo whose concentration is reduced so that it has the same $M_\star$ today as the black line. This galaxy inhabits a more massive halo yet has a lower star formation rate, compared to the black curves. The parameter $\alpha$ that characterises the halo concentration is held constant in these models.}
\label{fig:ikea}
\end{figure}

In this section, we make some simple predictions of the expected correlations to aid interpretation of the simulation results presented in the next section. The predictions are based on the \ikea\ model presented by \cite{Sharma2020}. This model posits that a galaxy's star formation rate $\dot M_\star \ (\equiv \rm{SFR})$ is proportional to the halo accretion rate, $\dot M_{\rm h}$, and can be written as
\begin{equation}
    \frac{1}{2}\,v^2_{\star}\,\dot M_{\star} = \frac{\kappa\,\alpha}{2}\omega_{\rm b}\,\dot M_{\rm h}\,\,v^2_{\rm h}\,;\quad v_{\rm h}^2\equiv \frac{{\rm G}M_{\rm h}}{R_{\rm h}\,.}
    \label{eq:ikea1}
\end{equation}
Here, $\kappa \approx 5/3$, $\omega_{\rm b} \equiv \Omega_{\rm b}/\Omega_{\rm m}$ is the fraction 
of the cosmological mass density in the form of baryons, and $v_\star$ is a characteristic velocity (see \citealp{Sharma2020} for details); 
$R_{\rm h}$ and $M_{\rm h}$ are the halo’s virial radius and mass, respectively, and  $v_{\rm h}^2$ measures the halo’s potential\footnote{In the original paper, the concentration parameter $\alpha$ is included in the definition of $v^2_{\rm h}$, see Eq.~(3). We don't do so here to better bring out the dependence on concentration.
This equation for the star formation rate superficially looks similar to that of \lq bathtub models\rq\ but is in fact significantly different in spirit as discussed briefly in \S \ref{sec:discussion} below and in more detail by \cite{Sharma2020}.}. The authors argue that $\dot M_\star$ tends to this secular star formation rate, provided that the star formation law is such that $\dot M_{\star}$ increases with the pressure in the star forming gas. If this is the case, then a halo with a deeper potential well, which has a larger value of $v^2_{\rm h}$, hosts a galaxy with a higher star formation rate. \cite{Sharma2020} show that \eagle\ galaxies follow Eq.~(\ref{eq:ikea1}) well.

The constant of proportionality between $\dot M_{\star}$ and $\dot M_{\rm h}$, depends on $v_{\star}$ and $\alpha$, measures of the efficiency of stellar feedback and of halo concentration, respectively. We will use
\begin{equation}
    \alpha\propto \frac{v^2_{\rm max}}{v^2_{\rm h}}\,,
    \label{eq:ikea2}
\end{equation}
where $v_{\rm max}$ is the maximum circular velocity, motivated by the paper of
\cite{Matthee17} who demonstrated that $M_\star$ correlates strongly with 
$v_{\rm max}$ at constant halo mass and redshift.

Since $v_{\rm h}^2\propto M_{\rm h}^{2/3}$ \cite[e.g.][]{Mo1998} and $\dot M_{\rm h}\propto M_{\rm h}$ \cite[e.g.][]{Correa2017}, we find that $\dot M_{\star}\propto M_{\rm h}^{5/3}\propto v^5_{\rm h}$ \citep{Sharma2020}. Therefore, \ikea\ predicts that $M_{\star}\propto M^{5/3}_{\rm h}$: this is the median $M_\star-M_{\rm h}$ relation in \ikea, which applies provided that all \ikea\ parameters (such as $\alpha$ and $v_\star$, for example) are redshift independent. For a halo of mass $M_{\rm h}=10^{12}~{\rm M}_\odot$ at $z=0$ and that follows the mean growth rate of haloes, as computed by \cite{Correa2017}, we plot the evolution of halo mass, stellar mass, and star formation rate as the black line in Fig.~\ref{fig:ikea}.

Scatter in the $M_\star-M_{\rm h}$ relation results because the rate at which a halo builds-up may differ from the median relation. To see how, consider haloes in a narrow halo mass range at $z=0$, say, and pick a halo that grew faster than average at early times (and hence slower at a later times): this is the case for the red dashed halo in Fig.~\ref{fig:ikea}. Such a halo typically has a concentration that is higher than average. How does this affect the stellar mass build-up? The relation between $v_{\rm h}$ and $M_{\rm h}$ depends explicitly on redshift, $v^2_{\rm h}\propto M^{2/3}_{\rm h}\,H^{2/3}(z)$ \cite[e.g.][]{Mo1998}, where $H(z)$ is the Hubble constant at redshift $z$. As a consequence, the galaxy in this halo forms stars at a {\em much} greater rate than average early on, and this more than compensates for its slower star formation rate at later times. As a consequence, at a given value of $M_{\rm h}$, higher concentration haloes that form earlier than average, host more massive galaxies than those that form later, and the star formation rate in those more massive galaxies is {\em lower} than average, as can be seen in the figure. The blue dotted line shows the corresponding evolution for a halo that is less concentrated and hence forms later: it hosts a galaxy with stellar mass lower than average which has a star formation rate that is higher than average. Important to note is that all three haloes have the same $z=0$ mass - yet different $M_\star$ and $\dot M_\star$. The correlation between concentration and $M_\star$ in \eagle\ galaxies was studied in detail by \cite{Matthee17}, who shows that concentration alone does not explain the full scatter. This is not surprising given that concentration does not uniquely characterize the evolution of halo mass.

However, rather than analysing galaxies at a given {\em halo} mass, typically galaxies are binned in {\em stellar mass}. The scatter in the $M_{\star}-M_{\rm h}$ relation and its dependence on formation path that we just discussed, can be used to predict correlations at given $M_\star$ as follows. Contrast the evolution of the dot-dashed red halo with that of the solid black halo in Fig.~\ref{fig:ikea}. This halo is more massive and therefore forms earlier. However, its concentration is lowered so that, according to the \ikea\ model, it hosts a galaxy with the same stellar mass as the halo in black. The star formation rate of the galaxy in the more massive but less concentrated halo is then {\em lower} at $z=0$ than that of the galaxy in the black halo. This is an example of how variations in formation path cause two galaxies with the same mass to inhabit haloes of different halo mass - with the galaxy in the more massive halo having a lower $\dot M_\star$. If we were brave, we might also predict that the star forming galaxy in the lower mass halo (black line), which is accreting high-angular momentum gas, is more likely to be discy, so that, at given $M_\star$, galaxies in the lower mass halo are star forming rotation-supported discs, whereas those in the more massive halo are forming stars at a lower rate and are more dispersion supported.

The correlations discussed in this section ultimately result from the scatter of haloes around the mean accretion history and how that itself correlates with halo concentration. However, the forming galaxy may itself affect its evolutionary path, for example in case the efficiency of stellar feedback changes maybe as a consequence of metallicity evolution. \cite{Kulier19} examines the scatter in the $M_\star-M_{\rm h}$ relation in \eagle, concluding that the feedback efficiency and the variations in the baryonic accretion rate play an important role. 

\cite{Sharma2020} show that the metallicity of the galaxy approaches a secular equilibrium value, $Z_{\rm eq}$, on the gas consumption time scale, $\tau_{\rm g}\equiv M_{\rm g}/\dot M_\star$, as described by their Eq.~(37):
\begin{equation}
\dot Z = \frac{y}{\tau_{\rm g}}\,\left(1-\frac{Z}{Z_{\rm eq}}\right)\,;\quad 
Z_{\rm eq} = y\,\frac{\kappa v^2_{\rm h}}{v^2_\star}\equiv \frac{y}{1+\eta}\,,
\label{eq:ikea3}
\end{equation}
where $y$ is the metal yield of the star-forming population and $\eta\equiv \dot M_{\rm outflow}/\dot M_\star$ is usually called the \lq mass-loading\rq\ factor. 
\cite{Almeida14} review similar relations in models that assume that galaxies are in a quasi-stationary phase where inflows and outflows balance the SFR.

The median MZR relation follows from recalling from the earlier discussion that $M_{\star}\propto v^5_{\rm h}$, yielding $Z_{\rm eq}\propto M^{2/5}_\star$ provided that $y$ and $v^2_\star$ - the efficiency of stellar feedback - remain constant\footnote{In his seminal paper, \cite{Larson72} neglected outflows and 
predicted that galaxies reach an asymptotic metallicity that is independent of their mass.}. At a {\em given} halo mass, the halo that forms earlier has higher $M_{\star}$ and hence also higher $Z$, we therefore expect a correlation between the residuals in $M_{\star}$ and $Z$ around their median value at given halo mass. Below we examine these correlations in \eagle\ further, including
correlations with the dynamics of the galaxies.

\section{The median $M_{\star}-Z$ relation (MZR) and its scatter in \eagle}
\label{sec:results}

\subsection{The stellar mass-metallicity relation at $z=0$}
\label{MZR_z0}

\begin{figure}
\centering
{\includegraphics[width=0.48\textwidth]{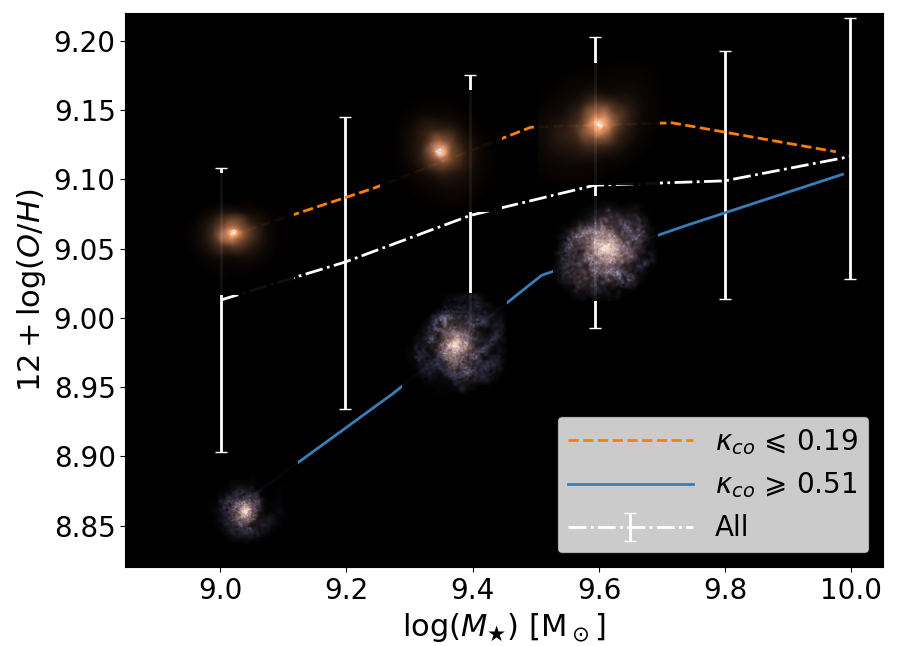}}
\caption{Median stellar mass-metallicity (MZR) relation of \eagle\ galaxies at $z=0$, with metallicity quantified by O/H. The MZR for the full sample is shown in {\em white dash-dotted lines} with 1$\sigma$ error bars.
The 20$^{\rm th}$ percentile of galaxies with the highest level of rotational support have $\kappa_{\rm co}\geqslant 0.51$ and are plotted in {\em solid blue}, the 20$^{\rm th}$ lowest have $\kappa_{\rm co}\leqslant 0.19$ and are plotted in {\em dashed orange}. Images of representative galaxies, created using {\sc{Py-SPHViewer}} \protect\citep{BenitezLlambay2015}, are shown along the sequences. High $\kappa_{\rm co}$ galaxies are blue and discy, low $\kappa_{\rm co}$ galaxies are red and spheroidal.}
\label{fig1}
\end{figure}

\begin{figure*}
\centering
\subfigure{\includegraphics[width=\textwidth]{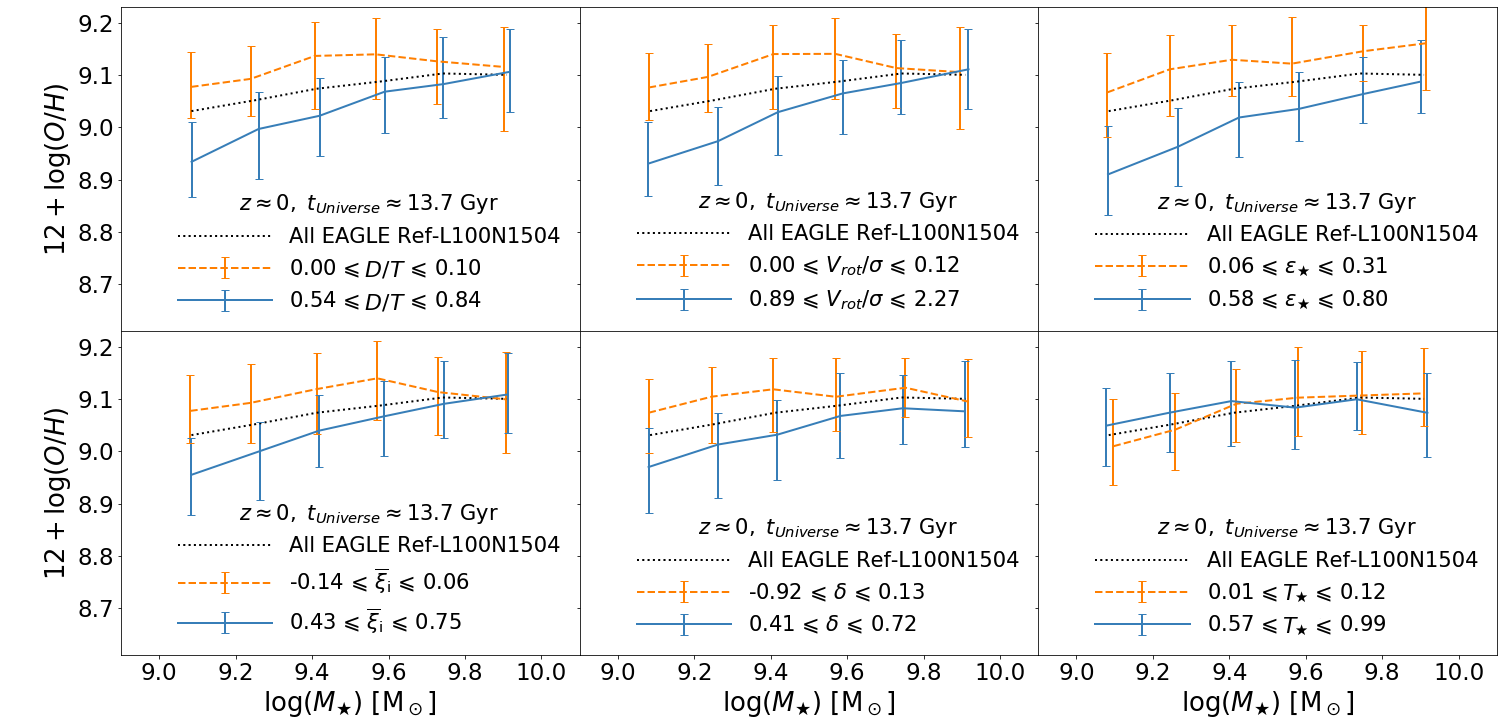}}
\caption{Median stellar mass-metallicity (MZR) of \eagle\ galaxies at $z=0$. The {\em black dotted line} in each panel shows the MZR for the full sample of galaxies. Different panels show the MZR when making different selections. Shown are: disc-to-total ratio ($D/T$, top left), rotation-to-dispersion velocity ratio ($V_{\rm rot}/\sigma$, top middle), stellar ellipticity ($\epsilon_\star$, top right), median orbital circularity ($\overline{\xi}_{\rm i}$, bottom left), velocity dispersion anisotropy ($\delta$, bottom middle), and triaxiality ($T_\star$, bottom right). As indicated in each panel, the {\em dashed orange} and {\em solid blue lines} represent the median relations for subsamples of $\approx 600$ galaxies, selected to be in the upper or lower 20$^{\rm th}$ percentile of the distribution; error bars enclose the corresponding 25$^{\rm th}$ and 75$^{\rm th}$ percentiles. Any criterion that separates galaxies between those that are rotationally supported (blue) versus dispersion supported (orange) results in also selecting predominantly low- versus high-$Z$ galaxies.}
\label{fig2}
\end{figure*}

We plot the median stellar mass-metallicity relation (MZR) at $z=0$ in Fig.~\ref{fig1} for the full sample of \eagle\ galaxies (white dash-dotted line), and the MZR for galaxies selected to have either higher- or lower levels of rotational support (blue solid and orange dashed lines are for galaxies with $\kappa_{\rm co}\geqslant 0.51$ and
$\kappa_{\rm co}\leqslant 0.19$, respectively). $Z$ increases with $M_\star$ for all galaxies, with the MZR in \eagle\ shallower than observed. 
As discussed by \cite{Schaye2015}, this is at least partially due to lack of numerical resolution, with a higher resolution \eagle\ run yielding a steeper relation. Incidentally, the observed value is close to that predicted by the \ikea\ model (see Fig.~14 of \citealt{Sharma2020}).

Galaxies with higher $\kappa_{\rm co}$ are typically bluer (higher sSFR), discy, more metal poor, and their MZR is steeper than average. In contrast, galaxies with lower levels of rotation are redder, more spheroidal, more metal rich, and have a shallower MZR than average. Clearly, the scatter in $Z$ at a given $M_\star$ is correlated with kinematics, morphology, and specific star formation rate of the galaxies. Given the different slopes of the MZR, eventually discy and spheroidal galaxies have similar $Z$, which occurs at $M_\star\sim 10^{10}~{\rm M}_\odot$, and at higher masses, discy galaxies have {\em higher} $Z$ than spheroidal galaxies of the same mass \citep[see also][]{Zenocratti2020}.

Parameters that quantify the morphology or kinematical properties of a galaxy are typically well correlated, as discussed by \cite{Thob2019}. These parameters also correlate
with $Z$, as shown in Fig.~\ref{fig2}. Selecting galaxies with a high level of rotational support or a discy morphology (blue solid lines), for example based on the disc-to-total ratio, $D/T$, or the rotation-to-dispersion velocity ratio, $V_{\rm rot}/\sigma$, etc., yields a MZR that is steeper and has lower amplitude. Selecting galaxies to be more spheroidal and with less rotational support (orange dashed lines), yields a MZR that is shallower and has higher amplitude. In addition to $D/T$ and $V_{\rm rot}/\sigma$, Fig.~\ref{fig2} shows this correlation when selecting by stellar ellipticity, $\epsilon_\star$,
median orbital circularities\footnote{The orbital circularity ${\xi}_{\rm i}$ of a given star particle is defined as the ratio of its angular momentum projected onto the rotation axis to the value it would have if the particle were on a circular orbit with the same binding energy. A positive (negative) value of this parameter corresponds to co- (counter-) rotation, and a value about $\approx 1$ corresponds to an almost circular orbit around the rotation axis. And, $\overline{\xi}_{\rm i}$ is defined as the mass-weighted mean of the orbital circularities of all star particles in the galaxy.}, $\overline{\xi}_{\rm i}$, 
or the velocity dispersion anisotropy\footnote{$\sigma_0$ and $\sigma_{\rm z}$, are, respectively, the velocity dispersion in the plane of the disc and in the direction of the rotation axis defined by all stars.}, $\delta=1-(\sigma_{\rm z}/\sigma_0)^2$. The triaxiality measure, $T_\star$, does not show a strong correlation with $Z$.

\begin{figure*}
\centering
\subfigure{\includegraphics[width=\textwidth]{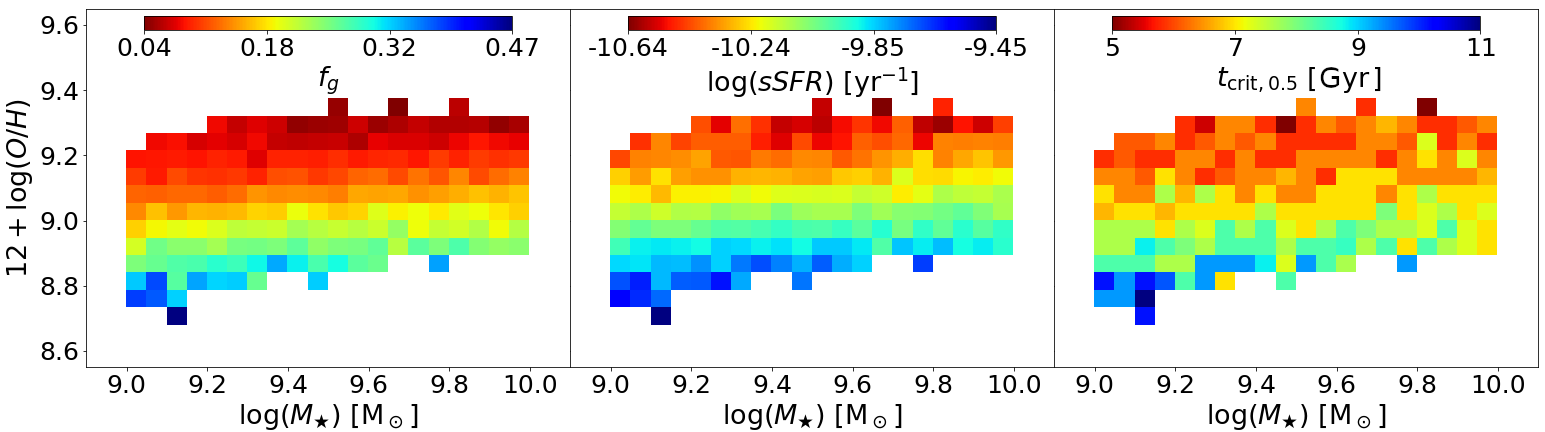}}
\caption{MZR plane for \eagle\ galaxies at $z=0$, colour-coded as function of different properties, namely as function of the fraction of star-forming gas ($f_{\rm g}$), specific star formation rate (sSFR), and cosmic time at which the galaxy formed half of its $z=0$ stellar mass ($t_{\rm crit,0.5}$, {\em left to right panels}, respectively). At a given value of $M_\star$, galaxies with high $f_{\rm g}$ have a high sSFR, formed relatively later, and have lower $Z$. }
\label{fig3}
\end{figure*}

\citet{DeRossi2017} examined how the metallicity $Z$ of star forming gas, quantified by its O/H, correlates with the fraction of star-forming gas ($f_{\rm g} \equiv M_{\rm{g}}/(M_{\rm{g}}+M_\star)$), sSFR and the mean stellar age, in \eagle\ Recal-L0025N0752 galaxies. They showed that galaxies with higher $f_{\rm g}$ tend to have higher sSFR, have younger stellar ages, and are more metal poor. We find similar correlations in \eagle\ Ref-L0100N1504, as shown in Fig.~\ref{fig3}, with the trend stronger at lower stellar mass.
The right panel of Fig.~\ref{fig3} shows that $Z$ also correlates with the mean stellar formation time $t_{\rm crit,0.5}$ - the value of cosmic time by which half the galaxy's current stellar mass has formed: at a fixed $M_\star$, younger galaxies (higher $t_{\rm crit,0.5}$) are more metal poor.

From Figs.~\ref{fig1}- \ref{fig3} we conclude that, at a given value of $M_{\star}$, $z=0$ \eagle\ galaxies with lower $Z$ have higher levels of rotational support (higher $\kappa_{\rm co}$), are more discy (higher $D/T$), are more gas rich (higher $f_{\rm g}$), and are younger (higher $t_{\rm crit,0.5}$). Our findings are consistent with those of \cite{DeRossi2017}; and in terms of the correlation between sSFR and morphology, with the analysis of \eagle\ galaxies by \cite{Correa2017}. These findings also agree with observations by \cite{Calvi2018}, who show that late-type (discy) galaxies have higher SFR compared to lenticular and elliptical galaxies of the same mass. The correlations between the kinematical parameter $\kappa_{\rm co}$ and the properties of the SF gas components of \eagle\ Ref-L0100N1504 galaxies were studied by \cite{Zenocratti2020}, who showed that, at $M_\star \la 10^{10} \ {\rm M}_{\sun}$, both $f_{\rm g}$ and sSFR increase with $\kappa_{\rm co}$, whereas the metallicity decreases.

\subsection{Origin of scatter in the MZR}
\label{sec:average_evolution}
 
\begin{figure*}
\centering
\includegraphics[width=\textwidth]{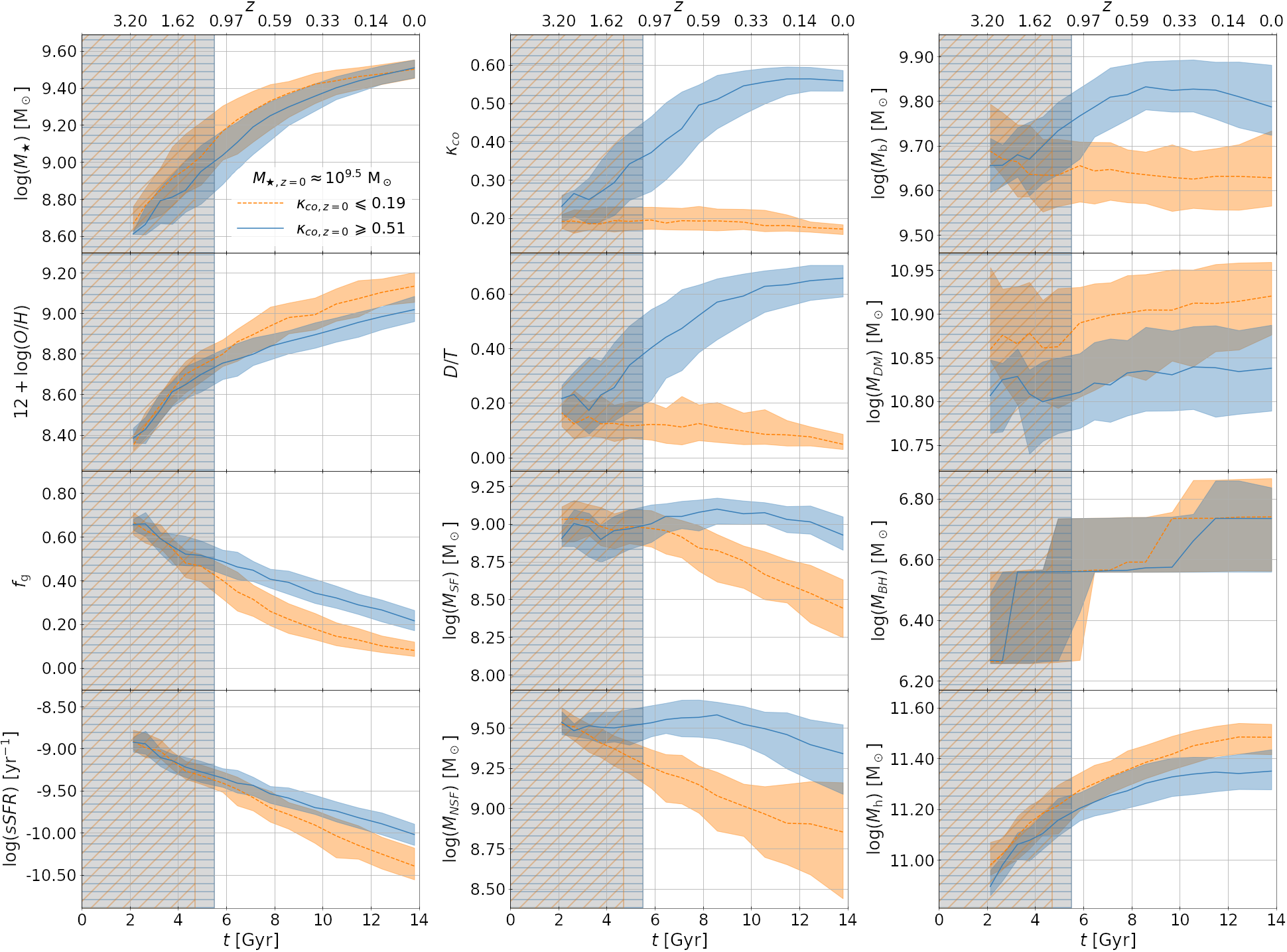}
\caption{Average evolution of properties of \eagle\ SF galaxies with $M_{\star} \approx 10^{9.5} \ \rm{M}_{\sun} $ at $z=0$, separating the sample into dispersion-supported (low $\kappa_{\rm co}$, orange dashed lines) and rotation-supported (high $\kappa_{\rm co}$, blue solid lines) galaxies. The colour-shaded regions around each curve enclose the corresponding $25^{\rm th}$ and $75^{\rm th}$ percentiles. The grey-shaded region with blue horizontal (orange slanted) solid lines enclose the cosmic times at which the average stellar mass of the rotation-supported (dispersion-supported) subsample is $M_\star < 10^9~\rm{M}_\odot$. Left panels, from top to bottom: evolution of stellar mass ($M_\star$), star-forming gas $O/H$ abundance, fraction of star-forming gas ($f_{\rm g}$), and specific star formation rate (sSFR). Middle panels, from top to bottom: evolution of rotational-to-total energy ratio ($\kappa_{\rm co}$), disc-to-total ratio ($D/T$), star-forming gas mass ($M_{\rm SF}$), and non-star-forming gas mass ($M_{\rm NSF}$). Right panels, from top to bottom: evolution of baryonic mass ($M_{\rm b}$), dark matter mass inside the galaxy ($M_{\rm DM}$), black hole mass ($M_{\rm{BH}}$), and mass of the halo that hosts the galaxy ($M_{\rm{h}}$). With the exception of $M_{\rm h}$, all quantities were calculated inside a spherical radius of 30 pkpc.}
\label{fig5}
\end{figure*}

\begin{figure*}
\centering
\subfigure{\includegraphics[width=0.5\textwidth]{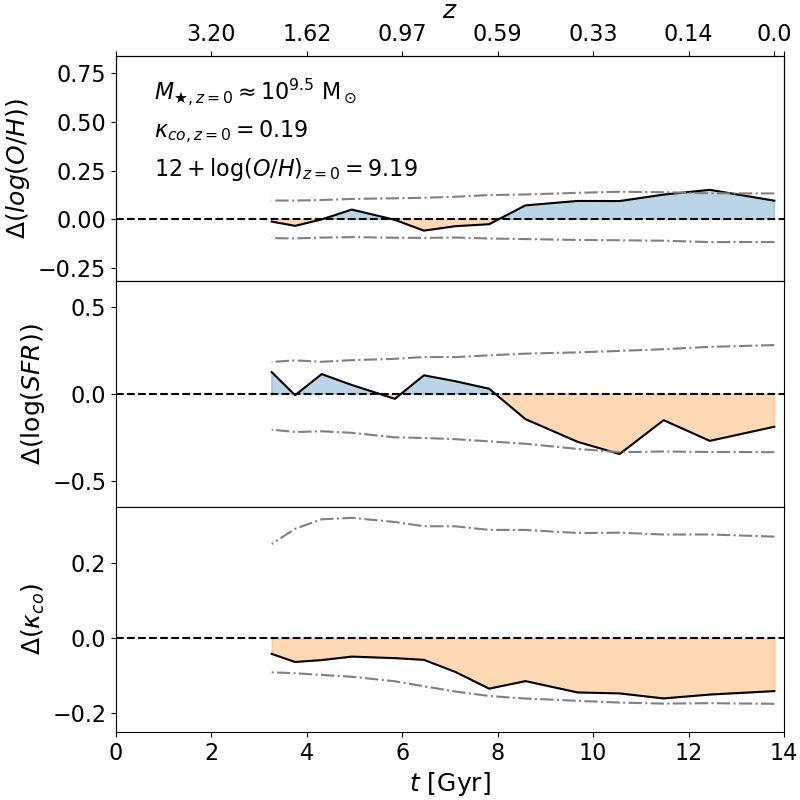}}
\subfigure{\includegraphics[width=0.5\textwidth]{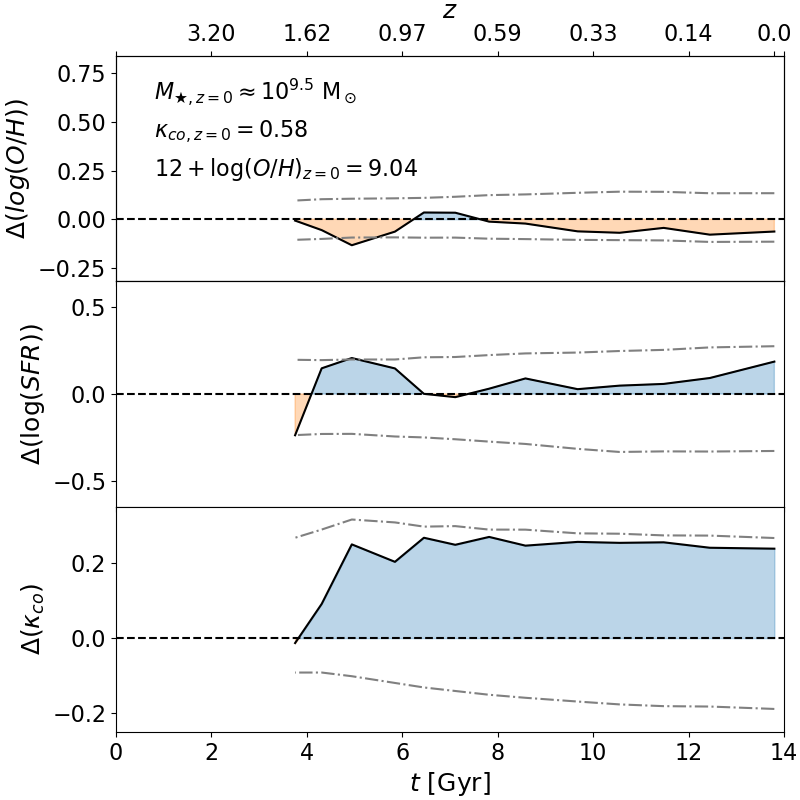}}
\caption{Evolution of the offsets of $\log({\rm O/H})$ ({\em upper panels}), $\log({\rm SFR})$ ({\em middle panels}), and $\kappa_{\rm co}$ ({\em lower panels}) from the median relation for two typical galaxies in the same $z=0$ mass bin ($M_{\star , z=0} \approx 10^{9.5}~\rm{M}_\odot$). The galaxy in the {\em left panel} is dispersion-supported at $z=0$, while the one in the {\em right panel} is rotationally supported. In each panel, the {\em blue} ({\em orange}) regions represent a positive (negative) offset from the median relation at that redshift. The {\em grey dashed-dotted lines} encompass the $25^{\rm{th}}$ and $75^{\rm{th}}$ percentiles of the underlying population
of galaxies with $M_{\star , z=0} \approx 10^{9.5}~\rm{M}_\odot$ at $z=0$.}
\label{fig6}
\end{figure*}

In this section we investigate whether scatter in the MZR, and the correlations discussed in the previous section between $Z$, $f_{\rm g}$, kinematics and stellar age, can be understood in terms of the formation history of the galaxy and/or its halo. To do so, we select galaxies in a narrow bin in stellar mass, 
$9.4 \leqslant \log(M_\star/\rm{M}_\odot) \leqslant 9.6$ at $z=0$. Next, we classify each galaxy based on the value of its rotational support at $z=0$, $\kappa_{\rm co}$, as either a rotationally supported system ($\kappa_{\rm co} \geqslant 0.51$; hereafter \lq RSS\rq\ galaxies) or a dispersion supported system ($\kappa_{\rm co} \leqslant 0.19$; hereafter \lq DSS\rq\ galaxies). Each sample consists of around 100 objects, and we trace their main progenitors back in time. We plot the evolution of several physical parameters of both subsamples in Fig.~\ref{fig5}, including $M_\star$, $Z$ and sSFR, as well as halo mass{\footnote{We take as halo mass $M_h \equiv M_{200}$ the total mass within the radius $R_{200}$, which is the physical radius within which the mean internal density is 200 times the critical density of the Universe, centred on the dark matter particle of the corresponding FoF halo with the minimum gravitational potential.}}, $M_h$, and black hole mass, $M_{\rm BH}$. Feedback from accretion onto the black hole does not strongly affect these galaxies, and we will not consider $M_{\rm BH}$ further. The evolution of the median and the scatter of the RSS and DSS samples are plotted in blue solid and orange dashed lines, respectively. We obtain similar trends when varying the choice of $M_\star$ at $z=0$ provided it is $\leqslant 10^{10}{\rm M}_\odot$.

By construction, the blue and orange samples overlap in stellar mass at $z=0$.
We also recognize the trends discussed before, where at a given value of $M_\star$, RSS galaxies have higher $\dot M_\star$ and lower $Z$; they also inhabit haloes with lower $M_h$. The blue and orange samples separate at $z>0$, and it is striking that even at redshift $z\sim 1$, differences in properties between the two 
samples are already in place. This was already pointed out by \cite{Kulier19}: properties of the progenitors at $z=1$ correlate strongly with scatter in stellar mass at given $M_h$ at $z=0$.

At early times, redshift $z\geqslant 1$, both galaxy samples have very similar sSFR, mass in star forming gas, $M_{\rm SF}$, mass in non-star forming gas, $M_{\rm NSF}$, and metallicity. However, there is already a significant difference in the level of rotational support, $\kappa_{\rm co}$ and $D/T$ ratio: DSS progenitors are more dispersion-supported and less discy at $z\sim 1$ than RSS progenitors. They also have higher $M_\star$ and halo mass, $M_h$; the latter means that the DSS halo started to form earlier. Also notable is that the baryonic mass of the DSS progenitors is significantly lower, even at $z=1$. At later times, both samples keep forming stars. However, the star forming gas fraction in the DSS progenitors falls much more rapidly with time than that of the RSS progenitors. This results in a significant decrease in $M_{\rm SF}$, an increase in $Z$, and the DSS progenitors' sSFR falls below that of the RSS progenitor. The drop in $M_{\rm SF}$ and $M_{\rm NSF}$ for the DSS progenitors is especially striking, given that
their dark matter mass does continue to increase: DSS galaxies 
have a low baryon accretion rate and their baryon mass remains practically constant. Their stellar mass increases and their SF gas mass decreases, so the SF gas is transformed into stars steadily, and there is no significant SF gas supply from accretion. The NSF gas also decreases, because it cools down (there is no relevant feedback to heat the gas or to prevent it from cooling down) and transforms into SF gas, somewhat alleviating the decrease in $M_{\rm SF}$ as function of time. In contrast, the baryonic mass, $M_{\rm b}$, increases significantly in RSS galaxies:
RSS galaxies keep accreting gas, their $M_{\rm SF}$ remains approximately constant, and a similar behaviour is found for $M_{\rm NSF}$, $\kappa_{\rm co}$ increases rapidly and so does $D/T$. It is worth mentioning that the increase in the halo mass for both subsamples implies that the host haloes accrete both dark matter and baryonic matter but this newly accreted baryonic material does not necessarily accrete onto the {\em galaxy}. In other words, an increase in $M_{\rm h}$ does not necessarily lead to gas accretion onto the central galaxy: this is what happens for DSS galaxies. In addition to this, the baryon fraction of the accreting material can itself be lower in the case of DSS progenitors.

The trends seen in Fig.~\ref{fig5} tally with the findings of \cite{Trayford19}, who studied the emergence of the Hubble sequence in \eagle. They find that at all $z$, stars tend to form in discy structures, but morphological transformations are common. Therefore, galaxies that have significant star formation tend to be or become discy, and those discs become increasingly dynamically cold at later times: this is the evolutionary path
of the RSS sample. If a galaxy has low star formation, then morphological transformations will tend to destroy any discs: this is the sequence of DSS galaxies.

The findings by \cite{Trayford19}, combined with the fact that we selected samples by requiring them to have the same $M_\star$ at $z=0$, goes some way in explaining the trends that we described. Indeed, selecting a galaxy to have low $\kappa_{\rm co}$ today picks out those galaxies which currently have low sSFR. Comparing them with galaxies with higher $\kappa_{\rm co}$ and hence higher sSFR at given $M_\star$, requires that the DSS galaxy formed earlier. This is most easily accomplished by increasing its halo mass, since
more massive haloes assemble earlier.

The presence of scatter in the $M_\star-M_h$ relation is crucial for the above scenario to be viable. Indeed, were there no scatter, the halo masses of all galaxies selected to have the same $M_\star$ should of course be the same. Therefore, the $z=0$ DSS galaxy with the same $M_\star$ as the RSS galaxy can only inhabit a more massive halo, if there is significant scatter in the $M_\star-M_h$ relation. The cartoon picture painted in Section~\ref{sect:expections} captures to some extent the trends seen here: the DSS galaxy with low $\dot M_\star$ has to form earlier to have the same mass today as the RSS 
that has higher $\dot M_\star$. Hosting it in a more massive halo accomplishes the earlier formation, but that halo needs to have a low concentration, otherwise $\dot M_\star$ for the DSS would be too high. We examine these trends in some more detail by studying the evolution of individual galaxies.

\subsubsection{The detailed evolution of individual galaxies}
\label{individual_evolution}

We analyse the trends discussed in the previous section for two galaxies, picked to be DSS or RSS at $z=0$, that have $M_{\star} \approx 10^{9.5}~\rm{M}_\odot$ at $z=0$. Similar trends are obtained at different stellar masses $M_{\star} \leqslant 10^{10}~{\rm M}_\odot$. 
We plot in Fig.~\ref{fig6} the evolution of the offsets of these galaxies from the median relation at that $z$, for O/H (upper panels), SFR (middle panels), and $\kappa_{\rm co}$ (lower panels). In each panel, orange and blue regions depict periods during which the property of each galaxy progenitor is higher or lower than the median of the galaxy population within the same $M_\star$ bin as the galaxy progenitor at the corresponding $z$ (i.e. positive or negative offsets), respectively. Grey dashed-dotted lines indicate the associated $25^{\rm th}$ and $75^{\rm th}$ percentiles of the background population for the same mass bins. We verified that the evolution of these two particular galaxies is representative for the evolution of DSS and RSS galaxies, respectively. In addition, as we checked by following the corresponding merger tree, neither of these galaxies underwent a major merger{\footnote{We define the level of merger $L_{\rm m}$ to quantify the importance of a merger, as $L_{\rm m}=m_2/m_1$, being $m_1$ the mass of the main progenitor of a galaxy, and $m_2$ the sum of masses of all the galaxies whose descendant is the galaxy under consideration. If $L_{\rm m}>1$ at a given snapshot, then the galaxy underwent a merger event between the current and the previous snapshots. We consider that $L_{\rm m}>1.33$ represents a major merger.}}. We plot each galaxy once it is resolved by more than 300 star particles ($M_\star\geqslant 3.5\times 10^{8}{\rm M}_\odot$).

The most striking trend is in the level of rotational support as shown by $\kappa_{\rm co}$: the DSS progenitor lacks rotational support over the whole of the plotted evolutionary path, whereas the RSS progenitor is strongly rotationally supported at all times. As we select these galaxies based on their value of $\kappa_{\rm co}$ at $z=0$, this trend is maybe not that surprising. The differences in SFR between the galaxies becomes more notable below $z\sim 1$, with the RSS progenitor more actively forming stars, and the DSS progenitor having low SFR. This results in a {\em decrease} in $Z$ for the RSS galaxy, and an increase in $Z$ for the DSS galaxy. This may seem surprising at first, but of course we showed in the previous section that the mass in star forming gas behaves very differently between these classes: the low amount of metals synthesized by the DSS galaxy enriches far less star forming gas in those galaxies. As a result, $Z$ actually increases {\em more rapidly} in the galaxy with the {\em lower} star formation rate: metallicity and SFR anti-correlate for these two galaxies. In summary, Fig.~\ref{fig6} shows that, at a given $M_\star$, significant increases of SFR offsets can be related to positive variations of $\kappa_{\rm co}$ and negative variations of O/H with respect to the median behaviour. We note that in general, SFR and $Z$ correlate {\em positively}, since more massive galaxies tend to have higher SFR and $Z$ on average. The anti-correlation we discuss here is for galaxies {\em of a given mass}. Such an anti-correlation is observed as well \cite[e.g.][]{Stott2013, Salim2015}.

\subsubsection{The impact of galactic inflows and outflows}
\label{sec:average_inout}

\begin{figure}
\centering
\includegraphics[width=0.45\textwidth]{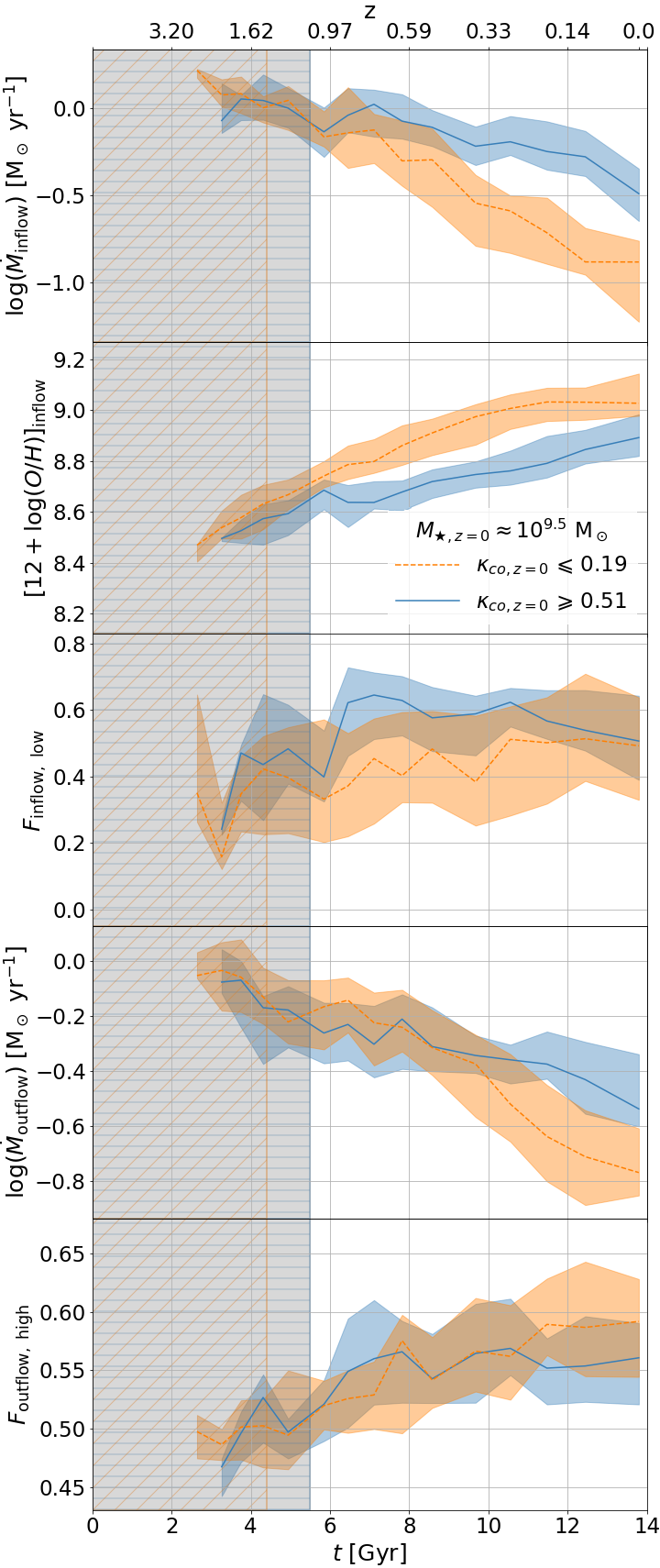}
\caption{Evolution of inflow and outflow rates for the progenitors of galaxies
with high (low) levels of rotational support, {\em blue} and {\em orange}, respectively;
the {\em shaded areas} enclose the 25$^{\rm th}$ and 75$^{\rm th}$ percentiles.
Galaxies are selected to have $M_\star\sim 10^{9.5}{\rm M}_\odot$ at $z=0$.
The {\em grey-shaded region} with blue horizontal (orange slanted) solid lines enclose the cosmic times at which the average stellar mass of the rotation-supported (dispersion-supported) subsample is $M_\star < 10^9~\rm{M}_\odot$. 
From top to bottom, the panels show the inflow rate of star-forming (SF) gas, $\dot{M}_{\rm inflow}$, its metallicity, $\log({\rm O/H})_{\rm inflow}$, 
the fraction of inflowing SF gas that has low metallicity, $F_{\rm inflow,low}$, 
the outflow rate of SF gas, $\dot{M}_{\rm outflow}$, and the fraction of outflowing SF gas with high metallicity, $F_{\rm outflow,high}$.
}
\label{fig7}
\end{figure}

\begin{figure*}
\centering
\subfigure{\includegraphics[width=0.33\textwidth]{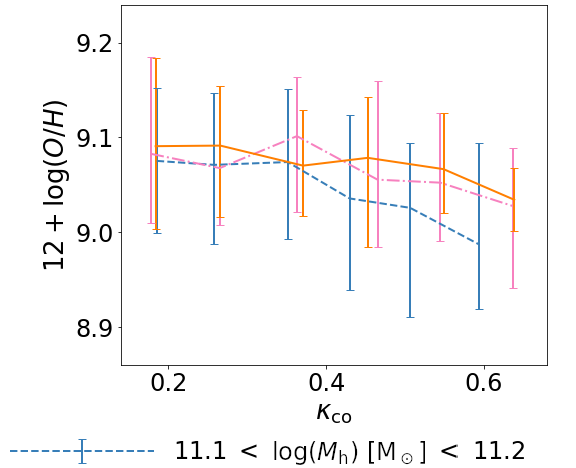}}
\subfigure{\includegraphics[width=0.33\textwidth]{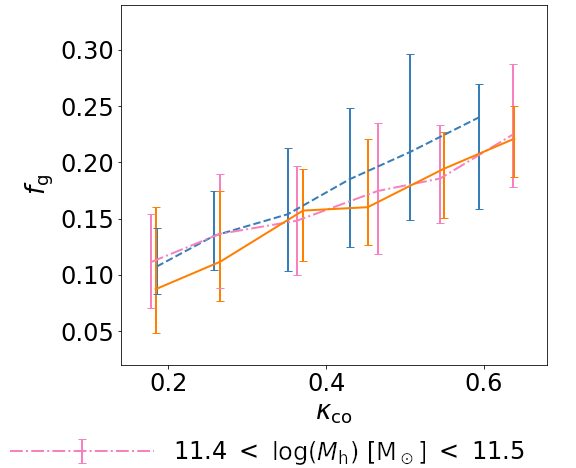}}
\subfigure{\includegraphics[width=0.33\textwidth]{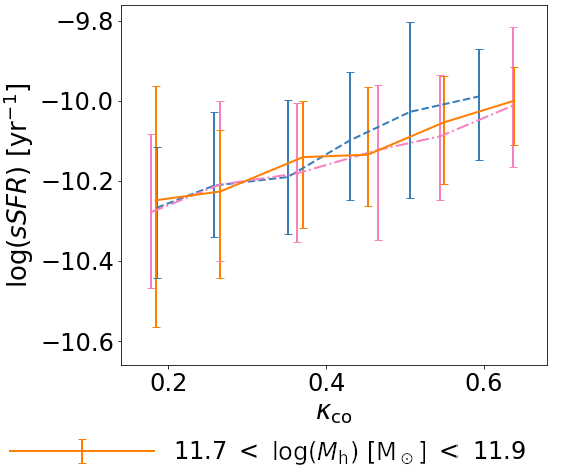}}
\caption{Correlation between O/H (left panel), the SF gas fraction $f_{\rm{g}}$ (middle panel), the specific star formation rate sSFR (right panel) and $\kappa_{\rm{co}}$,
the fraction of stellar kinetic energy invested in rotation, for central \eagle\ galaxies.
The {\em dashed blue}, {\em dot-dashed magenta} and {\em solid orange} lines correspond to galaxies with host halo mass in the range indicated in the legend. The curves represent the median relation, while error bars encompass the $25^{\rm{th}}$ and $75^{\rm{th}}$ percentiles. The main trends are the increase in $f_{\rm gas}$ and sSFR with $\kappa_{\rm co}$, and the increase in $Z$ with halo mass and decreasing $\kappa_{\rm co}$.}
\label{fig8}
\end{figure*}  

\begin{figure}
\centering
\subfigure{\includegraphics[width=0.4\textwidth]{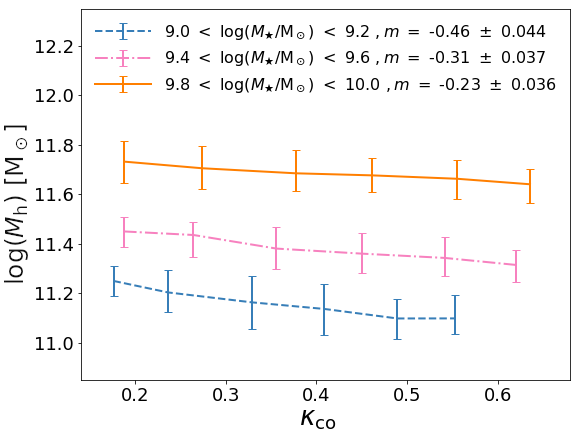}}
\par
\subfigure{\includegraphics[width=0.4\textwidth]{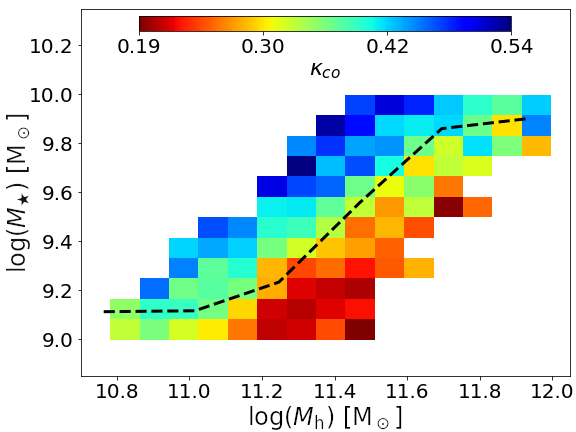}}
\par
\subfigure{\includegraphics[width=0.4\textwidth]{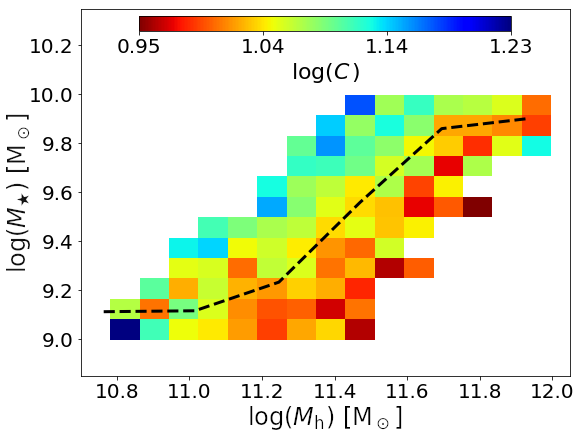}}
\caption{{\em Upper panel}: halo mass $M_{\rm h}$ as a function of $\kappa_{\rm{co}}$ of $z=0$ central \eagle\ galaxies, binned by stellar mass: low, intermediate and high $M_{\star}$ are shown by the {\em dashed blue}, {\em dot-dashed magenta}, and {\em orange solid line}, respectively. Curves denote the median relation, error bars encompass the $25^{\rm{th}}$ and $75^{\rm{th}}$ percentiles. The slope $m$ of the linear fit, $\log M_{\rm h}=\log M_{{\rm{h}}, {\rm ref}} + m\,\kappa_{\rm co}$ is reported in the panel. {\em Middle and lower panels}: the $M_{\rm h}$-$M_\star$ plane for these galaxies,
colour-coded according to $\kappa_{\rm co}$ and the concentration parameter
$C$ of the halo, respectively. The {\em black dashed line} is the median $M_{\rm h}$-$M_\star$ relation. At given $M_{\rm h}$, more concentrated haloes host more massive galaxies with higher $\kappa_{\rm co}$.}
\label{fig9} 
\end{figure}

The properties of inflowing and outflowing gas are computed by tracking
particles that cross a spherical aperture of 30~pkpc, as described in Section \ref{sec:accretion}. The Lagrangian nature of the simulation allows us to track individual particles, and makes it possible to compute inflow and outflow rates even in cases where the net amount of gas within 30~pkpc does not change. Here, we concentrate on properties of star-forming (SF) gas. Gas inside a 30~pkpc region that transforms from being not star-forming to star-forming between two snapshots is added to the inflow rate. Therefore, \lq inflow\rq\ refers to both SF gas that accretes onto a 30~pkpc radius and gas that becomes SF but was already inside a 30~pkpc radius. Similarly, \lq outflowing\rq\ SF gas refers to SF gas that leaves the 30~pkpc radius or gas that no longer satisfies \eagle's SF criterion. As mentioned in Section \ref{sec:accretion}, each inflowing and outflowing SF gas particle has its O/H abundance at a given snapshot, which can be compared with the overall abundance of the galaxy at the previous snapshot, so we can define the fraction of metal-poor inflowing gas, $F_{\rm{inflow,low}}$, as the mass fraction of inflowing SF gas whose O/H at a given snapshot is lower than the average O/H of the galaxy at the previous snapshot. A similar definition is used for the fraction of metal-rich outflowing gas, $F_{\rm{outflow,high}}$.

The results of this analysis are shown in Fig.~\ref{fig7}. As before, blue and orange colours refer to the progenitors of RSS and DSS galaxies, respectively. The main trends seen in this figure are as follows: before $z\sim 1$, there is little difference in the accretion properties of the two samples, apart from the slightly higher metallicity of the inflowing gas for DSS progenitors. At later times, there are more notable differences, in particular, the inflow rate $\dot M_{\rm{inflow}}$ of the DSS progenitors decrease much faster with time than that of the RSS progenitors, and the metallicity of their accreted gas {\em increases} faster. 
 Given the higher metallicity of the material accreted by DSS progenitors, we explore the role of ``galactic fountains'' for this sample. According to our results, the fraction of re-accreted particles that were previously ejected is lower than 10\% for galaxies in the DSS sample, so the contribution of ``galactic fountains'' to the galactic gas inflow is relatively small.
The outflow rate $\dot M_{\rm{outflow}}$ of the DSS progenitors is marginally higher than that of the RSS progenitors before $z\sim 0.5$, but then it is significantly below it. Thus, at recent times, galaxies in RSS seem to be affected by more significant SF gas inflows and more significant outflows, compared with DSS.

It is illuminating to compare the evolution of $\dot M_\star \ (\equiv \rm{SFR})$ with the inflow and outflow rates. From Fig.~\ref{fig5}, it can be estimated that for the RSS, $\dot M_\star\approx 0.5$ and 0.3~${\rm M}_\odot~{\rm yr}^{-1}$
at $z=1$ and $z=0$, respectively, as compared to inflow rates of $\dot M_{\rm inflow}=1$ and 0.3~${\rm M}_\odot~{\rm yr}^{-1}$ and outflow rates of $\dot M_{\rm outflow}=0.6$ and 0.25~${\rm M}_\odot~{\rm yr}^{-1}$. The ratio $\dot M_\star/\dot M_{\rm inflow}$ therefore changes from $0.5\to 1$: the $z=0$ galaxy is much more efficient at forming stars{\footnote{Although $M_{\rm{inflow}}$ also includes transformed gas, its fraction is small compared with the fraction of actually accreted gas, so the changes in efficiency are presumably more important than changes in the transformation rates.}}.
This is of course related to the evolution of the ratio $\dot M_{\rm outflow}/\dot M_\star$,
which changes from $1.2\to 0.8$: the $z=1$ galaxy drives a stronger outflow.

The higher inflow rate of the RSS progenitors results in a near constant reservoir of star forming gas, $M_{\rm SF}$, while their star formation rate drops (see Fig.~\ref{fig5}). Combined, this still leads to a gradual increase in $Z$ but at a rate more slowly than a galaxy with a more typical accretion rate. The latter grows along
the equilibrium sequence with $Z\approx Z_{\rm eq}$, given by Eq.~\ref{eq:ikea3}. 

In contrast, the inflow rate of the DSS progenitors drops precipitously (from $1\to 0.1~{\rm M}_\odot~{\rm yr}^{-1}$ at $z=1$ and $z=0$, respectively), much faster than their drop in $\dot M_\star$ (from $0.5\to 0.1~{\rm M}_\odot~{\rm yr}^{-1}$). This results in a more rapid increase in the metallicity of their ISM which we therefore describe by the \lq leaky box\rq\ model of galactic chemical evolution, rather than a closed box model. The leaky box model neglects inflows and assumes that the metallicity of the outflow is the same as that of the star forming gas. As a result, the metallicity increases as $M_{\rm SF}$ changes according to (see e.g. \citealp{Matteucci2003})

\begin{equation}
    \frac{\Delta Z}{Z_\odot} = -\frac{y}{Z_\odot\,(1+\eta)}\,\Delta\ln(M_{\rm SF})\,.
\end{equation}

Applying this to the DSS progenitor, we set $y=0.04$, read from Fig.~\ref{fig5} that $\dot M_\star \approx 0.1$ and that $M_{\rm SF}\approx 10^9$ and $10^{8.5}{\rm M}_\odot$ at $z=1$ and $z=0$, respectively, and read from Fig.~\ref{fig7} that $\dot M_{\rm{outflow}} \approx 0.16$, we have that $\eta \equiv \dot M_{\rm{outflow}}/\dot M_\star \approx 1.6$ and $\Delta\ln(M_{\rm SF}) \approx -1.1$ . We take $Z_\odot=0.0127$ for the \eagle\ model and the solar value of $A[{\rm O}]_\odot=12+\log({\rm O/H})_\odot=8.86$ from \citealt{Bergemann21}, where $A[{\rm O}]$ is the oxygen abundance by number. The oxygen abundance of the DSS galaxy at $z=0$, considering that $A[{\rm O}]_{z=1}\approx8.8$, can then be calculated as
\begin{eqnarray}
    A[{\rm O}]_{z=0} &=& A[{\rm O}]_{z=1} + \log_{10}\left(1+\frac
    {\Delta Z}{Z_\odot}10^{A[{\rm O}]_\odot-A[{\rm O}]_{z=1}}\right)\nonumber\\
    &\approx& 8.8+0.42=9.2\,,
\end{eqnarray}
in reasonable agreement with the values read from Fig.~\ref{fig5}.

With respect to the fraction of inflowing low metallicity SF gas, it can be seen from Fig.~\ref{fig7} that in RSS progenitors, $F_{\rm inflow,low}$ remains constant ($F_{\rm inflow,low} \approx 0.6$) between $z\approx 0.9$ and $z \approx 0.2$, and since then, it decreases reaching $F_{\rm inflow,low} \approx 0.5$ at $z = 0$: the recent and more significant SF gas inflows affecting RSS are mostly composed by relatively low-metallicity gas. On the other hand, in DSS progenitors $F_{\rm inflow,low}$ ranges between $\approx 0.4$ and $\approx 0.5$ at $0.9\leqslant z \leqslant 0$, so the SF gas inflows in those systems are not dominated by metal-poor material at recent times. These trends confirm that significant metal-poor gas inflows take place in RSS at late times, being this gas responsible for providing the fuel required for the significant increase of the disc mass of RSS galaxies at late times (see Fig.~\ref{fig5}). 

Finally, from Fig.~\ref{fig7}, it can be also seen that, on average, $F_{\rm outflow,high}$ ranges between $\approx 0.53$ at $z\la 1$ and $\approx 0.57$ at $z\approx 0.3$ for both populations, with more than 75\% of the galaxies showing $F_{\rm outflow,high} \ga 0.5$. Since $z\approx0.3$, $F_{\rm outflow,high}$ is, on average, somewhat higher for the DSS. But, in general, the outflowing SF gas is constituted by a relatively slightly more metal-enriched material compared with the SF gas metallicity of the host galaxies for both, RSS and DSS. It is worth noting that SF gas outflows can be associated to SN feedback caused by the star formation activity in our subsamples. Fig.~\ref{fig5} shows that galaxies in RSS tend to form most of their stellar component at later times, while, for DSS, most of the stellar assembly takes place earlier; this explains the higher outflows rates obtained for RSS towards lower $z$. 

If we compare the relative influence of inflows and outflows, we see that, for DSS,
$\dot{M}_{\rm inflow} \sim \dot{M}_{\rm outflow}$, roughly at all $z$, with inflows and outflows dominated in most cases by metal-enriched material compared with the metallicity of the corresponding host galaxies. On the other hand, for RSS, $\dot{M}_{\rm inflow} \gtrsim \dot{M}_{\rm outflow}$, with inflows dominated by metal-poor gas and outflows
composed by slightly more metal-enriched material. These different trends are in agreement with the weaker metallicity evolution obtained for RSS in Fig.~\ref{fig5} and explain why galaxies in DSS (RSS) tend to be located above (below) the median MZR, confirming our proposed scenario regarding the relevant role of metal-poor gas inflows on the determination of the metal-enrichment histories of rotation-supported SF galaxies with $M_\star \leqslant 10^{10}~\rm{M}_\odot$.

\subsection{Correlating the MZR with the properties of host dark matter haloes}

We investigate this correlation in Figs.~\ref{fig8} and \ref{fig9}. 
The first figure shows that $Z$ increases with $M_h$, as expected, but decreases with
$\kappa_{\rm co}$. The panels to the right show that galaxies with higher $\kappa_{\rm co}$ have a higher gas fraction and sSFR. These correspond to star-forming galaxies for which $Z$ increases as the host galaxy converts accreting gas to stars. This is in contrast to low $\kappa_{\rm co}$ galaxies, which form stars by converting their reservoir of star-forming gas to stars, with little or no replenishment by low-$Z$ accreting gas, as discussed in the previous section.

It is worth noting that, although there is a trend for galaxies with higher $\kappa_{\rm co}$ to have a higher sSFR at given $M_h$, the trend is small compared to the scatter, less than 0.2~dex (Fig.~\ref{fig8}, right panel). The differences in sSFR at a given $M_{\star}$ that we discussed in Fig.~\ref{fig5} are much larger, $\sim 0.4$~dex. The large difference in sSFR for galaxies at a given $M_{\star}$ is because ({\em i}) they inhabit haloes of different mass but mostly because ({\em ii}) the accretion rate on these haloes is very different. These differences also affect $\kappa_{\rm co}$, and the result is a much stronger correlation between $\kappa_{\rm co}$ and the sSFR. In other words:
at given $M_h$, higher $\kappa_{\rm co}$ galaxies have higher sSFR (Fig.~\ref{fig8}, right panel), but at given value of $M_{\star}$, that correlation is {\em much stronger}
(Fig.~\ref{fig5}). 

How the different behaviour of these galaxies relate to their halo is investigated in more detail in Fig.~\ref{fig9}. The two lower panels show that, at a given $M_h$, haloes that are more concentrated (those with a higher value of $C$, the concentration parameter of the NFW-fit to the dark matter profile) host a more massive galaxy that has
higher $\kappa_{\rm co}$. Conversely, at a given $M_{\star}$, the galaxy with higher $\kappa_{\rm co}$ has high concentration and inhabits a {\em lower} mass halo: these are the highly star-forming galaxies that convert low-$Z$ accreted gas into stars.
At that same value of $M_{\star}$, the galaxy with lower $\kappa_{\rm co}$ has lower concentration and inhabits a {\em more massive} halo: these are the DSS systems that are using up their reservoir of gas because of the lack of accretion of the halo.

In summary: at a given $M_{\star}$, the galaxy with higher sSFR is converting mostly accreted low-$Z$ gas into stars, and inhabits a concentrated, lower mass halo; the galaxy with lower sSFR is using-up its gas reservoir, and inhabits a lower concentration but more massive halo. This tallies with the cartoon picture of Section~\ref{sect:expections}.

\section{Comparison with previous work}
\label{sec:discussion}

We have shown that both the metallicity and the morphology of galaxies at given $z=0$ stellar mass $M_\star \lesssim 10^{10}~\rm{M}_\odot$, are strongly affected
by the presence or absence of $z\leqslant 1$ inflows of metal-poor gas in
the \eagle\ Ref-L0100N1504\ simulation. When a galaxy accretes gas, it forms stars
in a rotationally-supported disc, and those stars enrich the ISM, increasing its metallicity. On the other hand, if the gas accretion rate onto the galaxy is small, its metallicity increases faster and its amount of rotational support is smaller. All galaxy properties studied in this work were caclulated inside a
sphere with a radius of 30~pkpc, that is centered on the most bound particle
(lowest value of the gravitational potential), and accretion onto a galaxy 
results from both cosmological accretion and from accretion of recycled gas.
We find that cosmological accretion dominates in the galaxies that we studied,
with the recycled fraction $\approx 10\%$.

\citet{Sanchezalmeida2018} showed that, at a given stellar mass, central, star-forming \eagle\ galaxies that are physically larger have more metal poor and younger stellar populations. They argue that recent accretion of gas is the main reason for both correlations. Our results are consistent with these claims, in the sense that, at fixed $M_\star$, galaxies in our sample with lower metallicities have higher star formation activity, and hence they tend to have younger stellar components (see also, \citealt{DeRossi2017}). Taking into account that such systems exhibit higher levels of rotational support, and, as we showed in this work, they underwent recent accretion of metal-poor gas, disc-dominated, metal-poor galaxies in our sample tend to be more extended than their dispersion-supported, metal-enriched counterparts, in agreement with observations.

Analysing correlations between residuals of the MZR, the star formation rate, and the inflow and outflow rates of central \eagle\ galaxies, \citet{VanLoon2021} concluded that higher inflow rates correlate with lower metallicity. We agree with this conclusion, showing additionally that
the galaxies with higher inflow rate tend to be more rotationally supported.
Our findings also tally with those of \citet{DeLucia2020}, derived from predictions of the GAlaxy Evolution and Assembly ({\sc{GAEA}}) semi-analytic model. They show that an increase in the cold gas supply due to cosmological accretion leads to a decrease in metallicity.

\cite{Torrey2018} showed that galaxies in the {\sc{illustrisTNG}} simulations \citep{Pillepich2018} exhibit an anti-correlation between $Z$ and SFR, similar to the anti-correlation we described in Section~\ref{individual_evolution}. In their simulations, SFR and $Z$ oscillate about a median relation. They find
that the time scale associated with these oscillations is similar for both $Z$ and 
SF in their simulations, and claim that this is crucial for the emergence of a fundamental metallicity relation. \cite{Torrey2018} quote a value of $\sim 1.5$~Gyr for the variability time-scale at $z=0$. This value is comparable to the dynamical time-scale of haloes, $\tau_d$. In addition, the variability time-scale evolves with $z$, tracking the evolution of $\tau_d$, leading to the conclusion that the trends in SFR and $Z$ are driven by halo evolution. It is interesting to note that we do not seem to detect evolution in SFR and $Z$ on such a short time-scale. Nevertheless, long-term gas accretion can occur during successive shorter episodes of metal-poor gas inflow, each of which could generate moderate variations in gas fraction, metallicity and SFR. When averaged over several such episodes, the net effect is a decrease in metallicity, an increase in the gas fraction, and an increase in the star formation rate. When star formation is triggered after a single gas inflow event, stellar kinematics can be subject to minor variations in general, and, when averaging the behaviour over longer periods of time, the accumulated effects of successive episodes of gas accretion contributes to the formation of a galaxy disc. Interestingly we {\em do} find that SFR and $Z$ are affected by the evolution of the host halo. Given that the cosmological evolution of haloes is presumably nearly identical between \eagle\ and {\sc{illustris/illustrisTNG}}, the most obvious difference between the two sets of simulations is in the implementation of stellar feedback, and in particular, its effects on outflows.

In the \ikea\ model of \citet{Sharma2020}, the star formation rate $\dot{M}_\star$ does not depend on the gas mass, but it depends on the halo's accretion rate and potential well (see Eq.~(\ref{eq:ikea1})), and the star formation is regulated by feedback. This is a significant difference between \ikea\ and \lq gas-regulator\rq\ models (see Section 4 of \citealp{Sharma2020} for details). For instance, the `minimum bathtub' model (e.g. \citealp{Bouche2010, Dekel2014}) has very similar ingredients, however the star formation rate in such models is determined by the gas mass through a star formation law. In contrast, \ikea's star formation rate (Eq.~\ref{eq:ikea1}) is determined by the inflow rate, without reference to the gas mass, star formation law, or outflow rate. The \ikea\ model predicts that galaxies with deeper potential well have higher metallicity. Our results are consistent with this prediction.

Our results are also consistent with observational results obtained for galaxies at $M_\star \lesssim 10^{10}~\rm{M}_\odot$.  For example, \citet{VanDeSande2018} found that
when comparing galaxies of a given (low) mass in the {\sc sami} survey, galaxies with a higher level of rotational support are both younger and more discy. \citet{Wang2020} analysed {\sc MaNGA} data and concluded that, at a given value of $M_{\star}$,
galaxies with the highest star formation rates are late spirals, while systems with the lowest SFRs are spheroidal or \lq fast-rotator\rq\ early type galaxies. \citet{Bellstedt2021} concluded that $z<0.006$ galaxies in the {\sc gama} survey
occupy a well defined plane in the three-dimensional mass-metallicity-SFR space, with early type (late type) galaxies dominating the MZR at high (low) metallicities at a given mass. In summary, $M_\star \lesssim 10^{10}~\rm{M}_\odot$ galaxies in observed samples exhibit a correlation between kinematics, age, metallicity, and morphology, such that, at a given value of $M_{\star}$, galaxies with a higher level of rotational support are younger, more metal-poor and discy. The \eagle\ galaxies show similar correlations, and in the simulation, they are a consequence of the higher than average accretion rate of low-$Z$ gas on such galaxies.

\citet{Zenocratti2020} showed that in more massive galaxies,
$M_\star>10^{10}~\rm{M}_\odot$, not investigated here, the correlation between $Z$ and kinematics inverts, so that more dispersion supported galaxies have {\em lower} metallicity at a given stellar mass. In addition, $O/H$ decreases with decreasing $\kappa_{\rm co}$ in such galaxies. \citet{DeRossi2017} suggested that this can be understood by the feedback from active galactic nuclei (AGN) in \eagle, which becomes
increasingly important in these more massive galaxies. We aim to investigate this in more detail in future work.

\section{Summary and conclusions}
\label{sec:summary}

We have used the \eagle\ cosmological hydrodynamical simulations \citep{Schaye2015, Crain2015} to study the origin of the  mass-metallicity-morphology-kinematics relation, reported for the first time by \citet{Zenocratti2020} in this simulation. These galaxy properties are correlated, so that at a given stellar mass, $M_\star \leqslant 10^{10}~\rm{M}_\odot$, galaxies with higher levels of rotational support tend to be more metal-poor than dispersion-supported galaxies. In this paper, we focused on central, star-forming galaxies with stellar mass in the range $10^9~{\rm{M}}_\odot \leqslant M_\star \leqslant 10^{10}~\rm{M}_\odot$, and studied the properties of the progenitors of samples of galaxies that were selected to be rotationally supported (RSS) or dispersion-supported (DSS) at $z=0$. Our main conclusions are as follows:

\noindent
\begin{enumerate}[label=(\roman*),wide]

\item DSS galaxies tend to be older, are more spheroidal in shape, and are more metal enriched compared to RSS galaxies of the same stellar mass. The presence of DSS and RSS galaxies contributes significantly to the scatter around the mean mass-metallicity relation (MZR, Fig.~\ref{fig1}).

\item DSS galaxies have a significantly lower specific star formation rate (sSFR) compared to RSS galaxies of the same mass. This introduces an anti-correlation between star formation rate and $Z$, where galaxies with a higher SFR are more metal poor at a given mass: the fundamental metallicity relation, FMR \cite[see e.g.][]{Curti2020}.
DSS galaxies are also older than RSS galaxies of the same mass (Fig.~\ref{fig3}). These trends are consistent with the previous analysis of \eagle\ galaxies by
\citealp{DeRossi2017}.

\item The accretion rate of metal poor gas onto RSS galaxies is significantly higher below $z\sim 1$ than onto DSS galaxies of the same $z=0$ stellar mass. The gas fraction and the reservoir of star-forming gas, $M_{\rm SF}$, in such galaxies is therefore much higher. In contrast, DSS galaxies have much lower $M_{\rm SF}$ (Fig.~\ref{fig5} and \ref{fig7}).

\item The different levels of accretion onto galaxies introduces an anti-correlation between SFR and $Z$ at a given mass. RSS galaxies form stars out of the accreting, low-$Z$ gas, resulting in a gentle increase of $Z$ with $M_{\star}$. In contrast, DSS galaxies form stars at a reduced level by consuming their reservoir of star forming gas. The enrichment of an even smaller amount of gas results in a more rapid increase in $Z$. In this sense, DSS galaxies are well-described by the \lq leaky-box\rq\ model of galactic chemical evolution model.

\item The continued accretion of gas onto RSS galaxies allows them to form stars with high-levels of rotational support in a disc. Without continued accretion, the morphology of DSS galaxies and their progenitors becomes spheroidal. The continued accretion of gas seems required to enable the galaxy to form as well as sustain a disc.

\end{enumerate}

The accretion properties of galaxies are closely related to the concentration and formation history of their dark matter haloes. 
\begin{enumerate}[resume,wide]

\item At a given halo mass, $M_h$, RSS galaxies are more massive, have higher SFR and are more metal poor. The haloes of these RSS galaxies are more concentrated than average (Fig.~\ref{fig9}).

\item At a given stellar mass, RSS galaxies inhabit lower mass haloes than DSS galaxies.
However, the accretion rate onto the lower mass haloes is higher and more gas rich. This inflow sustains the star formation in the RSS galaxies, whereas the much lower accretion rate onto the DSS galaxies leads to a much reduced SFR.

\end{enumerate}

In summary, we find strong correlations between the scatter in the $M_{\star}-Z$ and the kinematic, morphological and SFR properties of galaxies. The root cause is to a large extent the different accretion properties of galaxies, itself related by the build-up of its halo. When a galaxy is able to continually accrete low-$Z$ gas, it continues to form stars in an increasingly discy, rotationally supported structure, with $Z$ increasing with $M_{\star}$ - the same process that results in the mean $M_{\star}-Z$ relation. On the other hand, if the accretion rate of gas is significantly lower, the galaxy forms stars by consuming its gas reservoir, which results in a high-$Z$ galaxy with little rotational support. At a given stellar mass, the rotationally supported galaxy inhabits a lower-mass, concentrated dark matter halo, whereas the dispersion-supported system inhabits a more massive halo. The more massive halo formed earlier, which is also why the stars in the DSS galaxy tend to be older.

Ultimately these trends result from ({\em i}) scatter in the concentration of dark matter haloes of a given mass, where more concentrated haloes form earlier, ({\em ii}) bias in the build-up of haloes, where the progenitors of more massive haloes form earlier. 
The more massive halo of the DSS progenitor forms its galaxy earlier compared to the lower mass RSS progenitor. When selecting galaxies to have the same $z=0$ stellar mass, this then requires that the accretion rate and hence the SFR of the DSS progenitor is much reduced compared to the RSS galaxy at later times. It is this sequence of events that results in an old, metal rich galaxy with low SFR in the case of the more massive halo, and a younger, more metal poor and more highly star forming galaxy in the lower-mass halo.

\section*{Acknowledgements}
We thank the referee for a careful reading of our manuscript and very constructive remarks which improved the paper.
LJZ and MEDR acknowledge support from PICT-2015-3125 of ANPCyT, PIP 112-
201501-00447 of CONICET and UNLP G151 of UNLP (Argentina).
We acknowledge the Virgo Consortium
for making their simulation data available. The EAGLE
simulations were performed using the DiRAC-2 facility at Durham,
managed by the ICC, and the PRACE facility Curie based in France
at TGCC, CEA, Bruy\`{e}res-le-Ch\^{a}tel.
This work used the DiRAC@Durham facility managed by the Institute for
Computational Cosmology on behalf of the STFC DiRAC HPC Facility
(www.dirac.ac.uk). The equipment was funded by BEIS capital funding
via STFC capital grants ST/P002293/1, ST/R002371/1 and ST/S002502/1,
Durham University and STFC operations grant ST/R000832/1. DiRAC is
part of the National e-Infrastructure.

\section*{Data Availability}

The \eagle\ simulations are publicly available. Both halo/galaxy catalogues and particle data can be accessed and downloaded at \url{www.icc.dur.ac.uk/Eagle/} \citep{Schaye2015,Crain2015,Mcalpine2016}. To create the results shown in this publication, Python libraries were used (\textsc{AstroPy}, \textsc{NumPy}, \textsc{PyPlot}
, and \textsc{H5Py}), including the publicly available \textsc{read\_eagle} module (\url{https://github.com/jchelly/read_eagle}, \citealp{eagle2017}). Additional data and code directly related to this work are available on reasonable request from the corresponding author.




\bibliographystyle{mnras}
\bibliography{biblio_Zenocratti_MZR_morpho_EAGLE} 




\appendix

\section{Convergence test}
\label{convergence}

\begin{figure}
\includegraphics[width=0.48\textwidth]{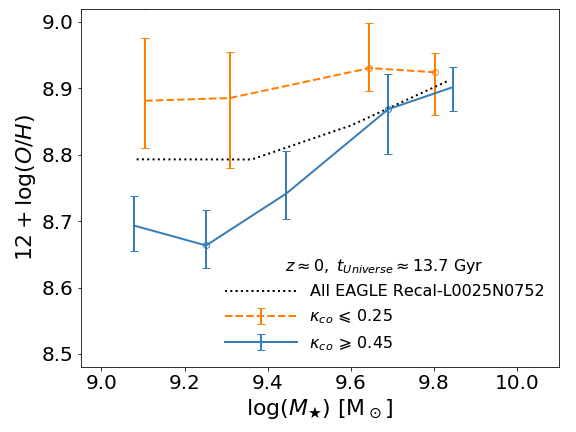}
\caption{Median MZR relation as function of the morpho-kinematical parameter $\kappa_{\rm co}$ (rotational-to-total energy ratio), at redshift $z=0$, for \eagle\ Recal-L0025N0752 simulated star-forming galaxies at $M_\star \leqslant 10^{10}~\rm{M}_\odot$. The dashed orange and solid blue lines depict the relations for galaxies with low and high rotational support (low and high values of $\kappa_{\rm co}$), respectively. The error bars denote the corresponding $25^{\rm th}$ and $75^{\rm th}$ percentiles. Open circles represent bins with less than 10 galaxies.}
\label{FigA1}
\end{figure}

\begin{figure}
\includegraphics[width=0.48\textwidth]{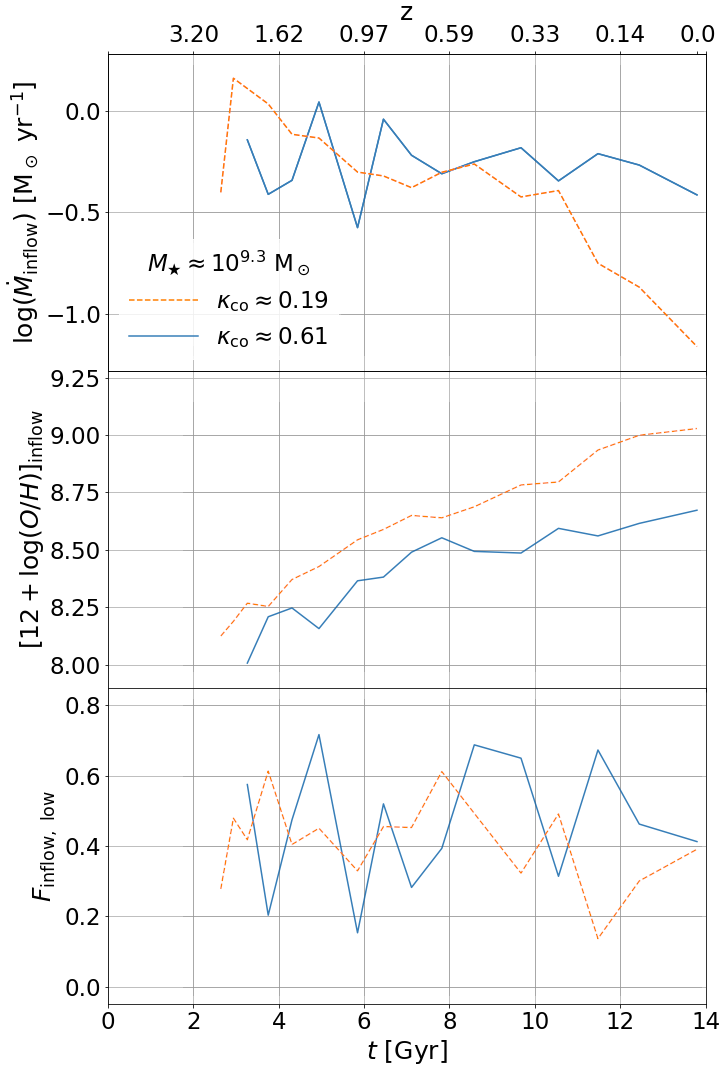}
\caption{
Analysis of SF gas inflows and outflows for two Recal-L0025N0752 galaxies.
The dashed orange lines correspond to a dispersion-supported galaxy (low $\kappa_{\rm co}$), while the solid blue ones depict a rotation-supported galaxy (high $\kappa_{\rm co}$), at $z=0$. From top to bottom, the panels show 
the SF gas inflow rate ($\dot{M}_{\rm inflow}$), metallicity of inflowing SF gas ($\log({\rm O/H})_{\rm inflow}$) and fraction of inflowing SF gas with low metallicity ($F_{\rm inflow,low}$).
}
\label{FigA2}
\end{figure}

So far, we have presented results from the intermediate resolution simulation
Ref-L0100N1504 because it provides a large galaxy sample that allowed a detailed statistical
analysis.
In this section, we test if our main findings are in agreement with
predictions from the high-resolution Recal-L0025N0752 simulation.
However, it is worth noting that the number of galaxies obtained from Recal-L0025N0752 by using
our selection criteria (Sec.~\ref{sec:galaxy_sample}) is significantly lower 
than that obtained from Ref-L0100N1504{\footnote{In Ref-L0100N1504 simulation, the sample obtained with our selection criteria consists of 4470 galaxies, while in Recal-L0025N0752 we found 105 objects using the same criteria.}}. Hence, the trends found for Recal-L0025N0752 are somewhat \lq noisier\rq\ and results from its statistical
analysis should be taken with care.

\begin{figure*}
\includegraphics[width=\textwidth]{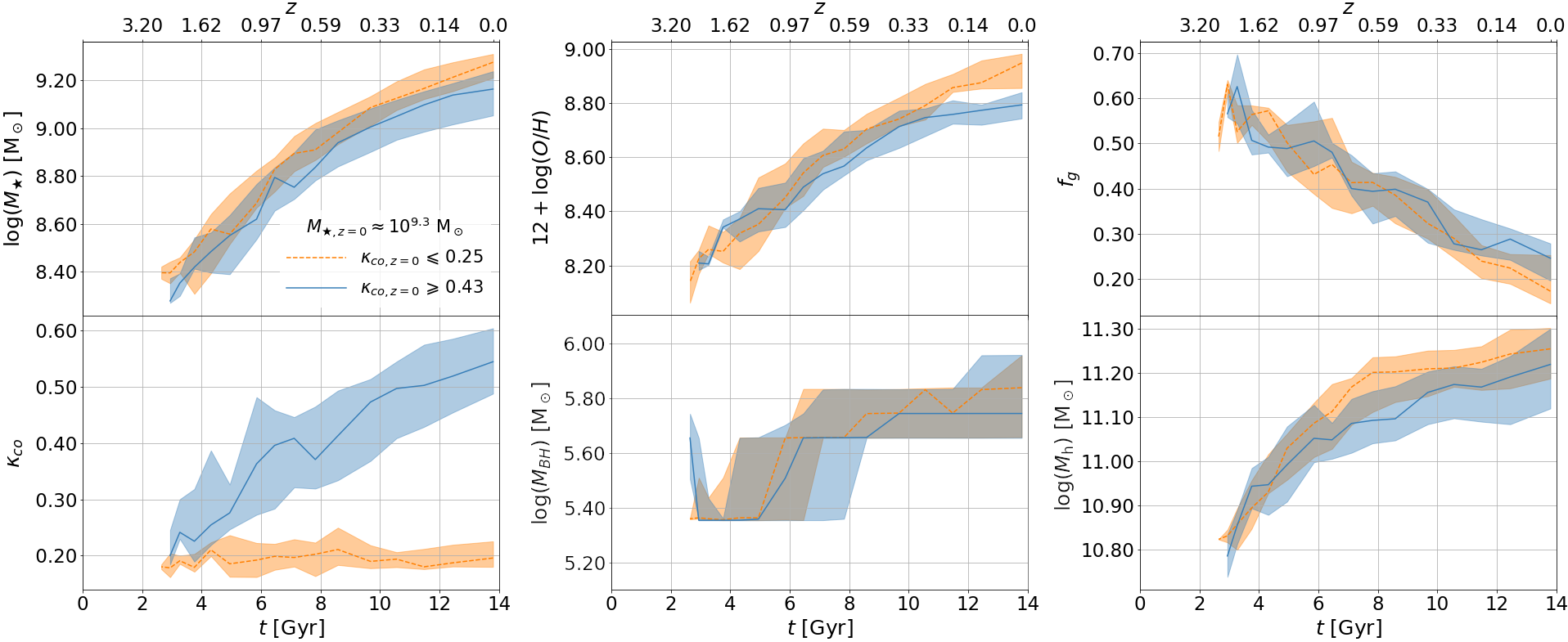}
\caption{Evolution of properties of low-mass ($M_{\star} \approx 10^{9.3} \ \rm{M}_{\sun} $) Recal-L0025N0752 galaxies, separating the sample into dispersion-supported (low $\kappa_{\rm co}$, orange dashed lines) and rotation-supported (high $\kappa_{\rm co}$, blue solid lines) galaxies. The colour-shaded regions around each curve enclose the corresponding $25^{\rm th}$ and $75^{\rm th}$ percentiles. From left to right, top panels show the evolution of stellar mass ($M_\star$), star-forming gas O/H, and fraction of star-forming gas ($f_{\rm g}$), while bottom panels show the evolution of rotational-to-total energy ratio ($\kappa_{\rm co}$), black hole-to-stellar mass ratio ($M_{\rm{BH}}/M_\star$), and baryonic-to-halo mass ratio ($M_{\rm b}/M_{\rm h}$).}
\label{FigA3}
\end{figure*}

In Fig.~\ref{FigA1}, we show the median MZR for Recal-L0025N0752 star-forming galaxies at $M_\star \leqslant 10^{10}~\rm{M}_\odot$, selected with the criteria described in Sec.~\ref{sec:galaxy_sample}. The figure shows the MZR for the complete sample (black dotted line) and for two extreme subsamples, separated
according to the kinematical parameter $\kappa_{\rm co}$. The dashed orange line represents the relation for dispersion-supported galaxies (low $\kappa_{\rm co}$), and the solid blue line corresponds to rotation-supported galaxies (high $\kappa_{\rm co}$). Even though Recal-L0025N0752 
predicts a steeper slope for the MZR (see, e.g, \citealt{DeRossi2017}), our findings regarding the O/H secondary dependences are similar
to those obtained from Ref-L0100N1504 simulation: at a fixed $M_\star$, rotation-supported systems have lower metallicities than dispersion-supported ones, being the difference in O/H roughly 0.2 dex at the lowest masses. As $M_\star$ increases, the differences in metallicity become smaller, being $\log({\rm O/H})$ practically constant for dispersion-supported galaxies, while it increases with $M_\star$ for rotation-supported systems. Since morpho-kinematical parameters correlate with each other, similar behaviours are found when studying the MZR as function of other parameters. 
Moreover, we also note that \citet{DeRossi2017} reported that secondary metallicity dependences on sSFR and $f_{\rm g}$ show similar
trends in Recal-L0025N0752 and Ref-L0100N1504.
Therefore, the secondary dependences of the MZR on the morpho-kinematics, sSFR and $f_{\rm g}$ of \eagle\ galaxies seem to be robust against resolution.

Given the difficulties to perform a robust statistical analysis with Recal-L0025N0752, Fig.~\ref{FigA2} shows a similar analysis of SF gas inflow to that in Fig.~\ref{fig7}, but for two {\em typical individual} galaxies extracted from Recal-L0025N0752. Dashed orange curves correspond to a dispersion-supported galaxy, and solid blue ones to a rotation-supported system. Comparing this figure with Fig.~\ref{fig7}, it can be seen that the behaviour of the different properties associated with inflows of SF gas is similar. At recent times ($z \lesssim 0.5$), the inflow rate of SF gas is higher for the rotation-supported system, with a more significant fraction of this inflowing SF gas being more metal-poor than the average O/H of the system.

Fig.~\ref{FigA3} shows the average evolution of some properties for two Recal-L0025N0752 subsamples of low-mass galaxies ($M_\star \approx 10^{9.3}~\rm{M}_\odot$), which are rotation- (solid blue curves) and dispersion-supported (dashed orange curves) at $z=0$. Although the differences between the evolution of some properties of these
two populations seem to be more modest for this simulation, the general trends shown here are the same as those shown, in Fig.~\ref{fig5}, for Ref-L0100N1504. In particular, we note that, given the lower number of galaxies in Recal-L0025N0752 simulation, we had to applied {\em weaker} constraints on $\kappa_{\rm co}$ when separating the subsamples. 
This last issue could generate the weaker differences between the evolution of Recal-L0025N0752 subsamples compared with those obtained, in Fig.~\ref{fig5}, for Ref-L0100N1504. In addition, in spite of our weaker constraints, the number of selected galaxies in Recal-L0025N0752 is still very low to perform a robust statistical analysis: in Fig.~\ref{fig5}, there are roughly 100 galaxies in each subsample of Ref-L0100N1504 simulation, while in Recal-L0025N0752, each subsample consists of around 15 galaxies.
Hence, results in Fig.~\ref{FigA3} should be taken with care.


\bsp	
\label{lastpage}
\end{document}